\documentclass[superscriptaddress,nofootinbib]{revtex4}
\pdfoutput=1
\usepackage{graphicx,epsfig}
\usepackage{float}
\usepackage[caption = false]{subfig}

\usepackage{amsmath}
\usepackage {amssymb}

\newcommand{\be}{\begin{eqnarray}}
\newcommand{\ee}{\end{eqnarray}}
\newcommand{\bea}{\begin{eqnarray}}
\newcommand{\nn}{\nonumber}
\newcommand{\eea}{\end{eqnarray}}

\newcommand{\fr}{\frac}
\def\de{\partial}

\makeindex

\begin{document}

\title{Spontaneous Momentum Dissipation and Coexistence of Phases in Holographic  Horndeski  Theory}
\author{George Filios}
\email{georgios.filios@uis.no} \affiliation{Department of Mathematics and Physics, University of Stavanger, 4036 Stavanger, Norway}
\author{P. A. Gonz\'{a}lez}
\email{pablo.gonzalez@udp.cl}
 \affiliation{Facultad de
Ingenier\'{i}a y Ciencias, Universidad Diego Portales, Avenida Ej\'{e}rcito
Libertador 441, Casilla 298-V, Santiago, Chile.}
\author{Xiao-Mei Kuang}
\email{xmeikuang@yzu.edu.cn} \affiliation{Center for Gravitation and Cosmology,
College of Physical Science and Technology,
Yangzhou University, Yangzhou 225009, China.}
\author{Eleftherios Papantonopoulos}
\email{lpapa@central.ntua.gr} \affiliation{Physics Division,
National Technical University of Athens, 15780 Zografou Campus,
Athens, Greece.}
\author{Yerko V\'{a}squez}
\email{yvasquez@userena.cl}
\affiliation{Departamento de F\'{\i}sica y Astronom\'ia, Facultad de Ciencias, Universidad de La Serena,\\
Avenida Cisternas 1200, La Serena, Chile.}
\date{\today}

\vspace{0.5cm}

\begin{abstract}
\begin{center}{\bf Abstract}\end{center}

We discuss the possible phases dual to the AdS hairy black holes in Horndeski theory.  In the probe limit breaking the translational invariance,  we study the conductivity and we find a  non-trivial structure indicating a collective excitation of the charge carriers as a competing effect of momentum dissipation and the coupling of the scalar field to Einstein tensor. Going beyond the probe limit, we investigate the spontaneous breaking of translational invariance  near the critical temperature and discuss the stability of the theory. We consider the backreaction of the charged scalar field to the metric and we construct numerically  the hairy black hole solution.  To determine the dual phases of a hairy black hole, we  compute the conductivity. When the wave number of the scalar field is zero, the DC conductivity is divergent due to the conservation of translational invariance. For nonzero wave parameter with finite DC  conductivity, we find two phases in the dual theory.
For low temperatures and  for positive couplings, as  the temperature is lower, the DC conductivity increases therefore the dual theory is in metal phase, while if the coupling is negative we have the opposite behavior and it is dual to an insulating phase. We argue that this behavior of the coupling of the scalar field to Einstein tensor can be attributed to its role as an impurity parameter in the dual theory.
\end{abstract}

\maketitle

\newpage
\tableofcontents

\section{Introduction}

One of the first applications of gauge/gravity duality \cite{Maldacena:1997re,Gubser:1998bc,Witten:1998qj} was the building of a holographic superconductor and the study of its properties. The basic mechanism of generating a holographic superconductor is the  formation in the gravity sector of a hairy black hole  below a critical temperature and the generation of  a condensate of the charged
scalar field  through its coupling to a Maxwell field of
the gravity sector \cite{Gubser:2008px}. Then it was shown  \cite{Hartnoll:2008vx} that a
holographic superconductor can be built on a boundary of
the gravity sector first on the probe limit, where the charge of the scalar field is much smaller than the
charge of the black hole, and then beyond the probe limit, in which the  backreaction of the scalar field to the spacetime metric was considered
\cite{Hartnoll:2008kx}. Considering fluctuations of the vector
potential, the frequency dependent conductivity was calculated,
and it was shown that it develops a gap determined by the
condensate. The formation of the gap is due to  a pairing mechanism which is  due to strong bounding of Cooper pairs. This strong pairing mechanism resulted to $2 \Delta \approx 8.4 T_c$, where $\Delta$ is the condensation gap, which has to be compared to
 the BCS prediction  $2 \Delta \approx 3.54 T_c$, which is much lower in real materials, due to impurities (for a review see   \cite{Horowitz:2010gk,Cai:2015cya}).

 The presence of impurities plays an important role in the superconducting materials chancing the properties of superconductors \cite{Abrikosov}. A detailed investigation of the effects of paramagnetic impurities in superconductors was carried out in \cite{skalski}, while  holographic  impurities were discussed in \cite{Hartnoll:2008hs}. A model of a gravity dual of a gapless
superconductor was proposed in \cite{Koutsoumbas:2009pa}. Below a critical temperature it was shown that a black hole
solution~\cite{Martinez:2004nb,Martinez:2005di} acquired scalar hair, the DC conductivity was calculated and it was found that it  had a milder behaviour than
the dual superconductor in the case of a black hole of flat
horizon~\cite{Hartnoll:2008vx}  which  exhibited a clear
gap behaviour. This  behaviour  was attributed  to materials with paramagnetic
impurities as it was discussed in \cite{skalski}. Impurities effects were studied in \cite{Ishii:2012hw}, while impurities in the Kondo
 Model were considered in \cite{O'Bannon:2015gwa,Erdmenger:2015xpq}.

 A holographic model of a superconductor was discussed in \cite{Kuang:2016edj}.   In this model the gravity sector consists
of a Maxwell field and a charged scalar field which except its minimal coupling to gravity it is
also coupled kinetically to Einstein tensor. This term belongs to a general class of
scalar-tensor gravity theories resulting from the Horndeski
Lagrangian \cite{Horndeski:1974wa}. The main effect of the presence of this kinetic coupling
is that  gravity  influences strongly  the  propagation of the scalar field compared to a scalar field minimally coupled to gravity.
It allows to find hairy black hole solutions \cite{Kolyvaris:2011fk,Rinaldi:2012vy,Kolyvaris:2013zfa,Babichev:2013cya,Charmousis:2014zaa,Anabalon:2013oea,Cisterna:2014nua,Babichev:2015rva},
the gravitational collapse of a scalar field with a derivative coupling to curvature takes more time for a black hole formation  compared to the collapse of minimally coupled scalar field  \cite{Koutsoumbas:2015ekk} and  acts as  a friction term in the inflationary period of the cosmological
evolution  \cite{Amendola:1993uh,Sushkov:2009hk,germani,Heisenberg:2018vsk}. Moreover, it was found that at the end of
inflation, there is a suppression of heavy particle production because of the fast decrease of  kinetic
energy of the scalar field coupled to curvature \cite{Koutsoumbas:2013boa}.

This behaviour of the scalar was used in \cite{Kuang:2016edj}
to holographically simulate the effects of a high concentration of impurities in a material. If there are impurities in a superconductor then the pairing mechanism of forming Cooper pairs is
  less effective  because the quasiparticles are loosing energy because of strong
concentration of impurities. Then it was found that as the value of the derivative coupling is increased the critical temperature is decreasing while
the condensation gap  is decreasing faster than the temperature. Also it was found
that the condensation gap for large values of the derivative coupling is not proportional  to the frequency of the real part of conductivity
which is characteristic of a superconducting state with impurities. Also a holographic superconductor was constructed in  \cite{Lin:2015kjk} with a scalar field coupled to curvature, and the holographic entanglement entropy  of a Horndeski black hole was calculated in \cite{Caceres:2017lbr}.

The real materials composed of electrons, atoms and so on, except their possible impurities, are usually doped.
This has important consequences in the properties of these materials and one of these is that they do not possess spatial translational invariance, and so the momentum
in the inner structure of these systems is not conserved.  An example of such an
observable is the conductivity.  Under applied fields, the charge carriers
of these systems can accelerate indefinitely if there is no mechanism for
momentum dissipation, which leads the DC conductivity to be infinite. However, once the spatial translational invariance is
broken which makes the momentum dissipation possible, then the peak  at the origin
of the AC conductivity spreads out such that the DC conductivity is finite. Therefore, for the  holographic theories it is important to incorporate the
momentum dissipation. In the literature, there are  plenty of studies on the transport
properties of holographic theories which dissipate momentum. The simplest way is to explicitly break the
translational symmetry of the field theory state \cite{Kim:2015dna}. Other ways are by  coupling to impurities, introducing  a  large amount of neutral scalar fields \cite{Andrade:2013gsa,Andrade:2014xca}, a special gauge-axion \cite{Li:2018vrz}, or by breaking translation invariance by introducing a mass for the graviton field \cite{Vegh:2013sk}.

  In the holographic superconductor practically there  is  a superconducting state which has a nonvanishing
charge condensate, and a normal state which is a perfect conductor.
The consequence of this is that even in the normal phase the DC conductivity ($\omega = 0$) is infinite, inspite of the proximity effect in the interface between the superconductor and the normal phase \cite{Kuang:2014kha}.  This is a  consequence of
the translational invariance of the boundary field theory, because the charge carriers do not dissipate their momentum, and accelerate freely under an
applied external electric field. Therefore one is motivated to introduce momentum
dissipation into the holographic framework, breaking the translational invariance of
the dual field theory.

In \cite{Donos:2013eha} black hole solutions were constructed which were holographically dual to strongly coupled doped theories with explicitly
broken translation invariance.  The gravity theory was consisted of and  Einstein-Maxwell
theory coupled to a complex scalar field with a simple mass term.  Black  holes dual to metallic and to insulating phases  were constructed and a their properties were studied.  Charged black brane solutions with translation  symmetry
breaking were considered in \cite{Gouteraux:2016wxj} and the conductivity at low temperatures was studied.

A generalization of \cite{Andrade:2013gsa} introducing a scalar field coupled to curvature was presented in \cite{Jiang:2017imk}.
An Einstein-Maxwell theory  was studied in which except the scalar fields present in \cite{Andrade:2013gsa}, scalar fields coupled to   Einstein tensor were also introduced. The holographic DC conductivity of the dual field theory was studied and the effects of the momentum dissipation due to the presence of  the derivative coupling were analysed. In \cite{Baggioli:2017ojd} the thermoelectric DC conductivities of Horndeski holographic models with momentum dissipation was studied in connection to quantum chaos. It was found that the derivative coupling  represents a subleading contribution to the thermoelectric conductivities in the incoherent limit.

To explain the generation of FFLO states \cite{Fulde, Larkin} an interaction term between
the Einstein tensor and the scalar field is introduced in a  model \cite{Alsup:2012ap,Alsup:2012kr} with two
U(1) gauge fields and a scalar field coupled to
a charged AdS black hole. In the absence of an interaction of the
Einstein tensor with the scalar field, the system possesses
dominant homogeneous solutions for all allowed values of
the spin chemical potential. Then calculating the DC conductivity it was found that the system exhibits
a Drude peak as expected. However,
 in the presence of the interaction term, at low temperatures, the system is shown to
possess a critical temperature for a transition to a scalar
field with spatial modulation as opposed to the homogeneous solution.

In this work we spontaneously break the translational invariance in an Einstein-Maxwell-scalar gravity theory in which
the charged scalar field is also coupled kinetically to Einstein tensor and we  study the possible  phases generated on the dual boundary theory.
In the probe limit we calculated the conductivity and we found that depending on the parameter of the translational symmetry breaking and on the coupling of the scalar field to curvature we have on the dual theory a coexistence of phases on  the boundary theory. Then we go beyond the probe limit and considering the fully backreacted problem we constructed numerically a hairy black hole solution. To determine the  phases of the dual theory to the hairy black hole, we  compute the conductivity. When the wave parameter of the scalar is zero, the DC conductivity is divergent due to no mechanism of momentum relaxation, which is dual to ideal conductor. While for nonzero wave parameter with finite DC  conductivity, we found two phases in the dual theory. For low temperatures, we found that for positive couplings, as we lower the temperature the DC conductivity increases therefore the dual theory is in metal phase, while if the coupling is negative we have the opposite behavior and it is dual to an insulating phase. We attributed this behavior of the coupling of the scalar field to Einstein tensor to that this coupling is connected to the amount of impurities present in the theory. In the procedure, the system first is with the scalar field spacially dependent and the scalar potential has a constant chemical potential. Then we perturb the system around the critical temperature. We find that the scalar develops an $x$-dependence backreacted solution, which implies the spontaneously  generated inhomogeneous phase of the system.

The work is organized as follows. In Section \ref{sec-setup} we set up the problem. In Section \ref{sec-HS} we studied the probe limit of the theory. In Section \ref{sec-criticality} we discussed  the stability of the theory and calculated the critical temperature. In Section \ref{sec-backreaction} we calculated numerically the fully backreacted black hole solution and studied the DC conductivity and finally in Section \ref{sec-conc} are our conclusions.

\section{The field equations in Horndeski theory}\label{sec-setup}

The action in Horndeski theory, in which  a complex scalar field which except its coupling
to metric it is also coupled to Einstein tensor reads
\bea\label{EGBscalaraction}
   I=\int  d^4x\sqrt{-g}\left[ \frac{R-2\Lambda}{16\pi G}-\fr{1}{4}F_{\mu\nu}F^{\mu\nu}- \left(g^{\mu\nu}  (D_\mu\Psi)^\ast D_\nu\Psi + m^2|\Psi|^2+ \beta G^{\mu\nu} (D_\mu \Psi)^\ast D_\nu\Psi\right) \right]~,
\eea
where \be D_\mu = \nabla_\mu - i q A_\mu \ee
and $q$, $m$ are the charge and the mass of the scalar field and
$\beta$ is the coupling of the scalar field to Einstein tensor of
dimension length squared. For convenience we define
 \bea
\Phi_{\mu\nu} &\equiv& D_{\mu}\Psi (D_{\nu}\Psi)^*~,\\
\Phi &\equiv& g^{\mu\nu}\Phi_{\mu\nu}~,\\
C^{\mu\nu} &\equiv& g^{\mu\nu} + \beta G^{\mu\nu}~.
 \eea
Subsequently, the field equations resulting from the action
(\ref{EGBscalaraction}) are
 \bea G_{\mu\nu} +\Lambda g_{\mu\nu} = 8\pi  T_{\mu\nu}
\ , \ \ \ \ T_{\mu\nu} = T_{\mu\nu}^{(\Psi)} +
T_{\mu\nu}^{(EM)} + \beta\Theta_{\mu\nu}~, \label{einst}\eea where,
\bea
T_{\mu\nu}^{(\Psi)} & = &   \Phi_{\mu\nu} + \Phi_{\nu\mu} - g_{\mu\nu}(g^{ab}\Phi_{ab} + m^2 |\Psi|^2)~, \\
T_{\mu\nu}^{(EM)} & = & F_{\mu}^{\phantom{\mu} \alpha} F_{\nu
\alpha} - \fr{1}{4} g_{\mu\nu} F_{\alpha\beta}F^{\alpha\beta}~,
\eea and \bea
\Theta_{\mu\nu}  = & -& g_{\mu\nu} R^{ab}\Phi_{ab} + R_{\nu}^{\phantom{\nu}a}(\Phi_{\mu a} + \Phi_{a\mu}) + R_{\mu}^{\phantom{\mu}a} (\Phi_{a\nu} + \Phi_{\nu a})  - \fr{1}{2} R (\Phi_{\mu\nu} + \Phi_{\nu\mu}) \nn\\
& - & G_{\mu\nu}\Phi - \fr{1}{2}\nabla^a\nabla_\mu(\Phi_{a\nu} + \Phi_{\nu a}) - \fr{1}{2}\nabla^a\nabla_\nu(\Phi_{\mu a} + \Phi_{a\mu}) + \fr{1}{2}\Box (\Phi_{\mu\nu} + \Phi_{\mu\nu}) \nn \\
& + & \fr{1}{2}g_{\mu\nu} \nabla_a\nabla_b (\Phi^{ab} + \Phi^{ba})
+ \fr{1}{2}(\nabla_\mu\nabla_\nu + \nabla_\nu\nabla_\mu) \Phi -
g_{\mu\nu}\Box\Phi~,\label{theta} \eea
and the Klein-Gordon equation is
\be (\de_\mu-i q A_\mu) \left[
\sqrt{-g}C^{\mu\nu}(\de_\nu - i q A_\nu)\Psi\right] =
\sqrt{-g} m^2\Psi~, \label{glgord} \ee
while the Maxwell equations read \be \nabla_\nu F^{\mu\nu} +
C^{\mu\nu} \left[ 2 q^2 A_\nu |\Psi|^2 + i q
(\Psi^*\nabla_\nu\Psi- \Psi\nabla_\nu\Psi^*)\right]
=0~. \label{max}\ee
We note that the matter action in (\ref{EGBscalaraction}) has a $1/q^2$ in front,
so the backreaction of the matter fields on the metric is
suppressed when $q$ is large and the  limit $q\rightarrow\infty$ defines the probe limit. When $q$
goes to zero, we have neutral scalar field and it has no coupling with the Maxwell field.

\section{Signature of breaking the translation symmetry in the probe limit }
\label{sec-HS}

We firstly focus on the probe limit in the above setup, in which the Einstein equations
admit  the planar Schwarzschild AdS black hole solution
 \be\label{schwar}
 ds^2= -f(r) dt^2 + \frac{dr^2}{f(r)} + r^2 (dx^2 +
 dy^2)\qquad,\qquad
f(r)=r^2 - \frac{r_h^3}{r}~. \ee
 By setting $\Psi=\Psi(z)$ and $A=\phi(z)dt$ in \cite{Kuang:2016edj} a holographic superconductor was built in the probe limit in the background of the black hole \eqref{schwar}.

In this section we will break the translational invariance leading to momentum dissipation
and calculate the conductivity in the probe limit.
To this end, we consider  the ansatz for the matter fields as
\begin{equation}
\Psi(r,x)=\varphi(r) e^{- i\tau x},~~~~A_\mu=\phi(r) dt~,
\end{equation}
where $\tau$ indicates the strength of the breaking of the translational invariance.
Then under the metric (\ref{schwar}), the equations of motion for $\varphi(r)$ and $\phi(r)$ become
\begin{eqnarray}\label{psieq}
\left[1+\beta\left(\frac{f}{r^2}+\frac{f'}{r}\right)\right]\varphi^{\prime\prime} + \left[\frac{2}{r} + \frac{f^\prime}{f} + \beta\left (\frac{3f'}{r^2}+\frac{f'^{2}}{r f}+\frac{f''}{r} \right)\right]\varphi^\prime + \left[\frac{q^2\phi^2}{f^2}\left(1+\beta\left(\frac{f}{r^2}+\frac{f'}{r}\right)\right) - \frac{m^2}{f}\right]\varphi \nonumber\\
-\frac{\tau^2\varphi}{r^2 f}\left[1+\beta\left(\frac{f'}{r}+\frac{f''}{2}\right)\right]= 0 ~,
\end{eqnarray}
\begin{eqnarray} \label{phieq}
 \phi^{\prime\prime} + \frac{2}{r}\phi^\prime -\frac{\tau^2}{r^2 f}\phi- \frac{2q^2\varphi^2}{f}\left[1+\beta\left(\frac{f}{r^2}+\frac{f'}{r}\right)\right]\phi = 0~.
\end{eqnarray}
Because of the breaking of the translational invariance in both the scalar field and vector potential equations we have an extra term multiplied by $\tau$. We note that when  $\tau=0$ and $\beta\neq 0$, the system recovers our previous  work \cite{Kuang:2016edj}. And when $\tau=0$ and $\beta=0$, this model goes back to the minimal holographic superconductor of \cite{Hartnoll:2008vx}.

Near the boundary, the matter fields  behave as
\begin{equation}\label{eq-boundayM}
\varphi | _{r \rightarrow \infty}= \frac{\varphi_-}{ r^{\Delta_-}}+\frac{\varphi_+}{ r^{\Delta_+}},~~~\phi=\mu-\frac{\rho}{r}
\end{equation}
and  $\Delta$ is
\begin{equation}\label{eq-Delta}
\Delta_{\pm}=\frac{3}{2}\pm\frac{\sqrt{9+4m^2+27 \beta}}{2 \sqrt{1+3 \beta}}~,
\end{equation}
where $\mu$ and $\rho$ are interpreted as the chemical potential and charge density in the
dual field theory, respectively.

According to the AdS/CFT correspondence, either of the  coefficients $\varphi_{-}$ and
$\varphi_{+}$ may correspond to the source of an operator $\mathcal{O}$ dual to the scalar field with the other dual to  the vacuum expectation values
$\langle\mathcal{O}_{-}\rangle=\sqrt{2}\varphi_{-}e^{- i\tau x}$, $\langle\mathcal{O}_{+}\rangle=\sqrt{2}\varphi_{+}e^{- i\tau x}$, depending on the mass window and the choice of quantization.
In details, from \eqref{eq-Delta} , the condition $m^2\geq -\frac{9}{4}(1+3\beta)$ is needed for the relevance of the operator. When  $-\frac{5}{4}(1+3\beta)\geq m^2\geq -\frac{9}{4}(1+3\beta)$,  both $\varphi_{-}$ and $\varphi_{+}$ are normalizable and we can choose both modes as source while the other as  vacuum expectation values. When $m^2\geq -\frac{5}{4}(1+3\beta)$, only $\varphi_+$
is normalizable, so we can only treat $\varphi_+$ as vacuum expectation value of the operator while $\varphi_-$ as the source. Moreover, we shall give two comments about the dual theory. First,
 $\beta$  modifies the mass windows which are different from that in Einstein case \cite{Horowitz:2008bn}. It is obvious that $\beta=-1/3$ leads the dual theory to be ill-defined and the bulk theory at $\beta=-1/3$ may have no good holographic dual description in the boundary. Second, by redefining $m^2\to (1+3\beta)m^2$, the dual theory described above seems to reduce the theory dual to Einstein scalar theory. This question is worthy to further study because with the redefinition, the action \eqref{EGBscalaraction} from the bulk side does not apparently have similarity as the  Einstein-scalar theory.

\begin{figure}[h]
\center{
\includegraphics[scale=0.8]{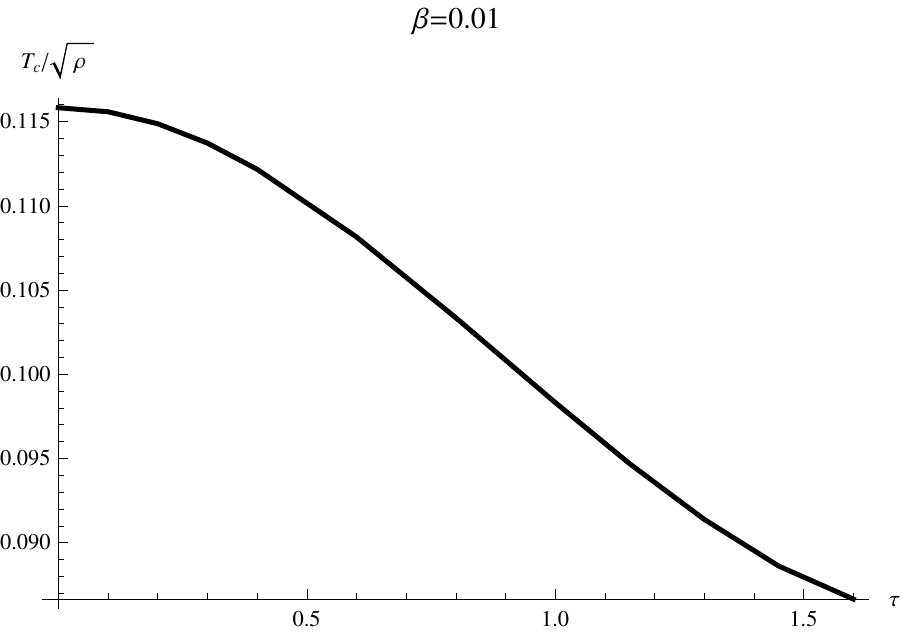}\hspace{0.5cm}
\includegraphics[scale=0.8]{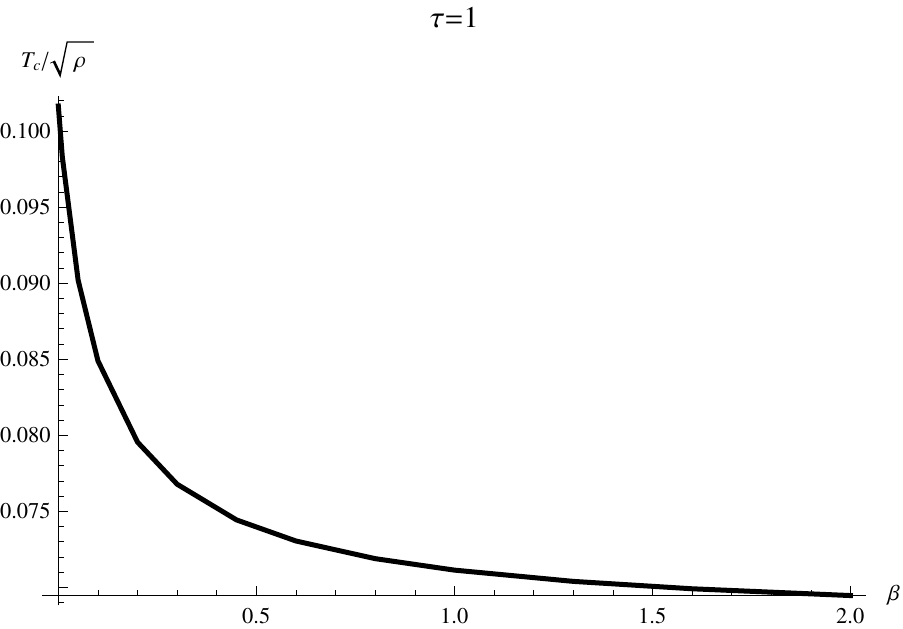}\\
\caption{\label{fig_Tc} The critical temperature of phase transition depending on $\tau$ (left) and $\beta$ (right). }}
\end{figure}
\begin{figure}[h]
\center{
\includegraphics[scale=0.8]{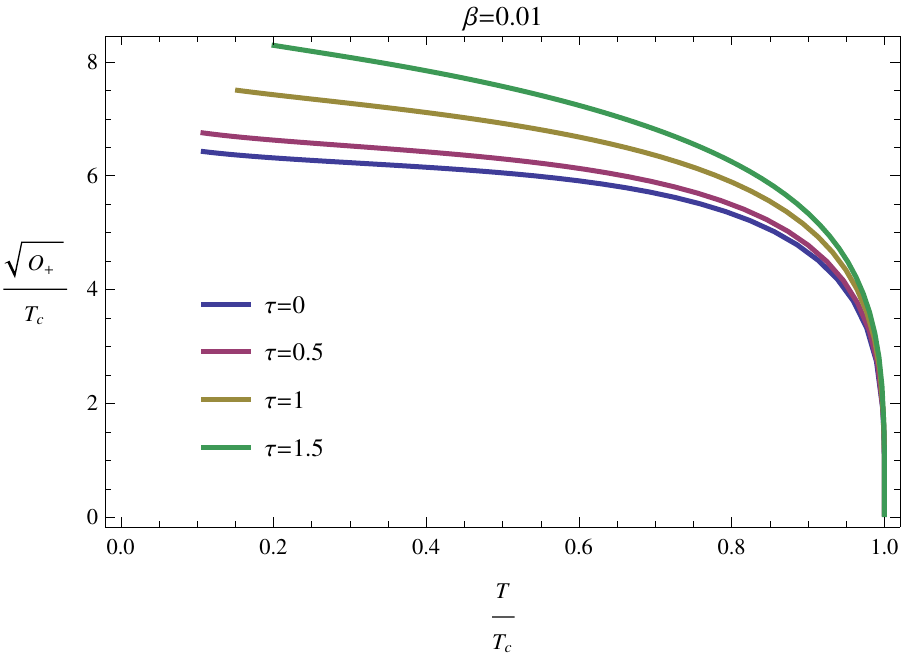}\hspace{0.5cm}
\caption{\label{fig_O} The strength of condensation at $x=0$ for fixed $\beta=0.01$.}}
\end{figure}

The equations (\ref{psieq})  and (\ref{phieq}) can be solved
numerically by doing integration from the horizon  to the
infinity by taking regular condition near the horizon. Then we extract the data near the boundary. As we decrease the temperature to a critical value $T_c$, there exists a phase transition from normal black to superconducting black hole. We choose the boundary condition $\varphi_-=0$. So the solutions  correspond to vanishing $\varphi_+$ for normal phase  and non-vanishing $\varphi_+$ for superconducting phase, respectively. We show the critical temperature $T_c$ in Fig~\ref{fig_Tc}. We can see that as the derivative coupling $\beta$ and the momentum dissipation parameter $\tau$ are increasing, the  critical temperature is decreasing, indicating that the system is harder to enter the superconducting phase.

This behaviour can also be seen in Fig.~\ref{fig_O} which shows the  condensation gap at the position $x=0$. For a fixed value of the derivative coupling $\beta$, as the temperature of the system decreases, the strength of the condensation gap is enhanced as the dissipation parameter $\tau$ is increased.

To compute the conductivity in the dual CFT as a function of
frequency, we need to solve the Maxwell equation for fluctuations of the vector
potential $A_x$.   The Maxwell equation at
zero spatial momentum and with a time dependence of the form $e^{-i \omega t}$ gives
\begin{equation}\label{eq:Axeq}
A_x''+\frac{f'}{f}A_x' +\left[\frac{\omega^2}{f^2}-\frac{2\varphi^2}{f}\left(1+\beta\left(\frac{f''}{2}+\frac{f'}{r}\right)\right)\right]A_x -\frac{2q\tau \varphi ^2}{r^2}\left(1+\beta\left(\frac{f'}{r}+\frac{f''}{2}\right)\right)=0~.
\end{equation}

We will solve the perturbed Maxwell equation with ingoing wave boundary conditions at the horizon, i.e.,
 $A_x  \propto f^{-i\omega/3r_h}$. The asymptotic behaviour of the Maxwell field at large radius  is  $A_x = A_x^{(0)} + \frac{A_x^{(1)}}{r} + \cdots$. Then,
according to AdS/CFT dictionary,  the dual source and expectation
value for the current are given by $ A_x = A_x^{(0)} $ and $ \langle J_x \rangle = A_x^{(1)}$, respectively.
Thus, the conductivity is read as
\begin{equation}
\sigma(\omega)= - \frac{i A_x^{(1)}}{\omega A_x^{(0)}}~.
\end{equation}
\begin{figure}[h]
{\includegraphics[scale=0.5]{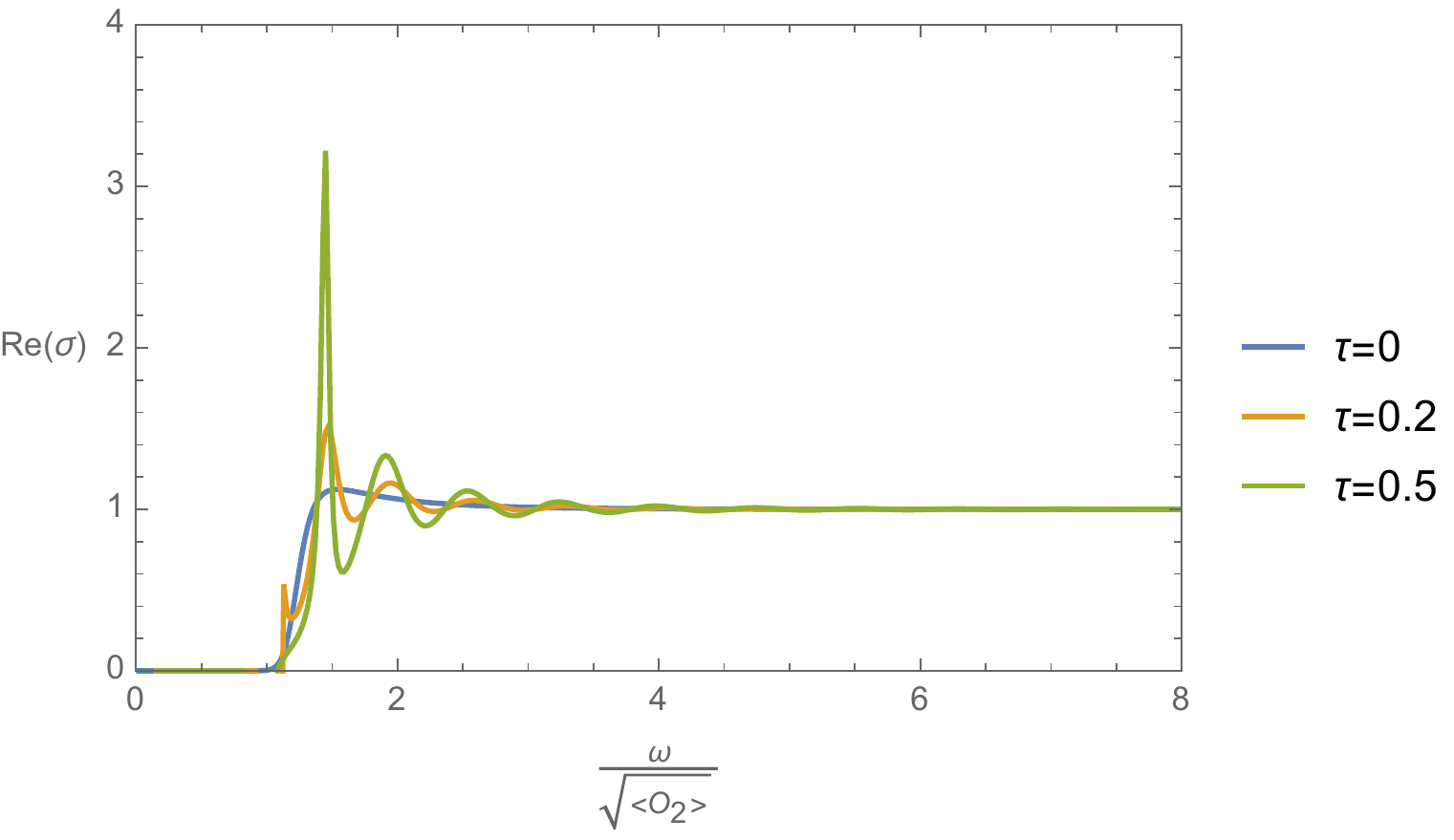}}
{\includegraphics[scale=0.5]{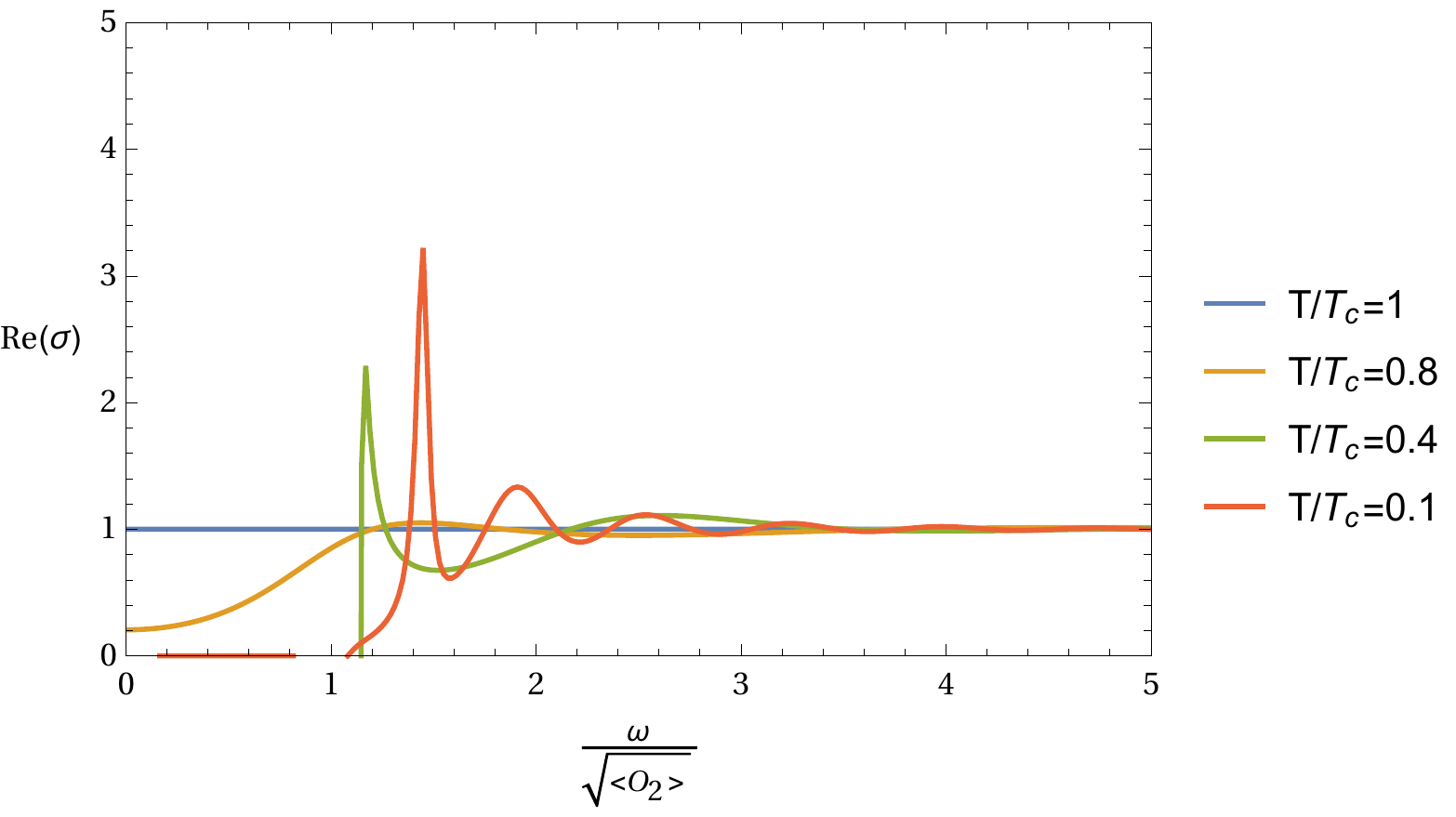}}
{\includegraphics[scale=0.5]{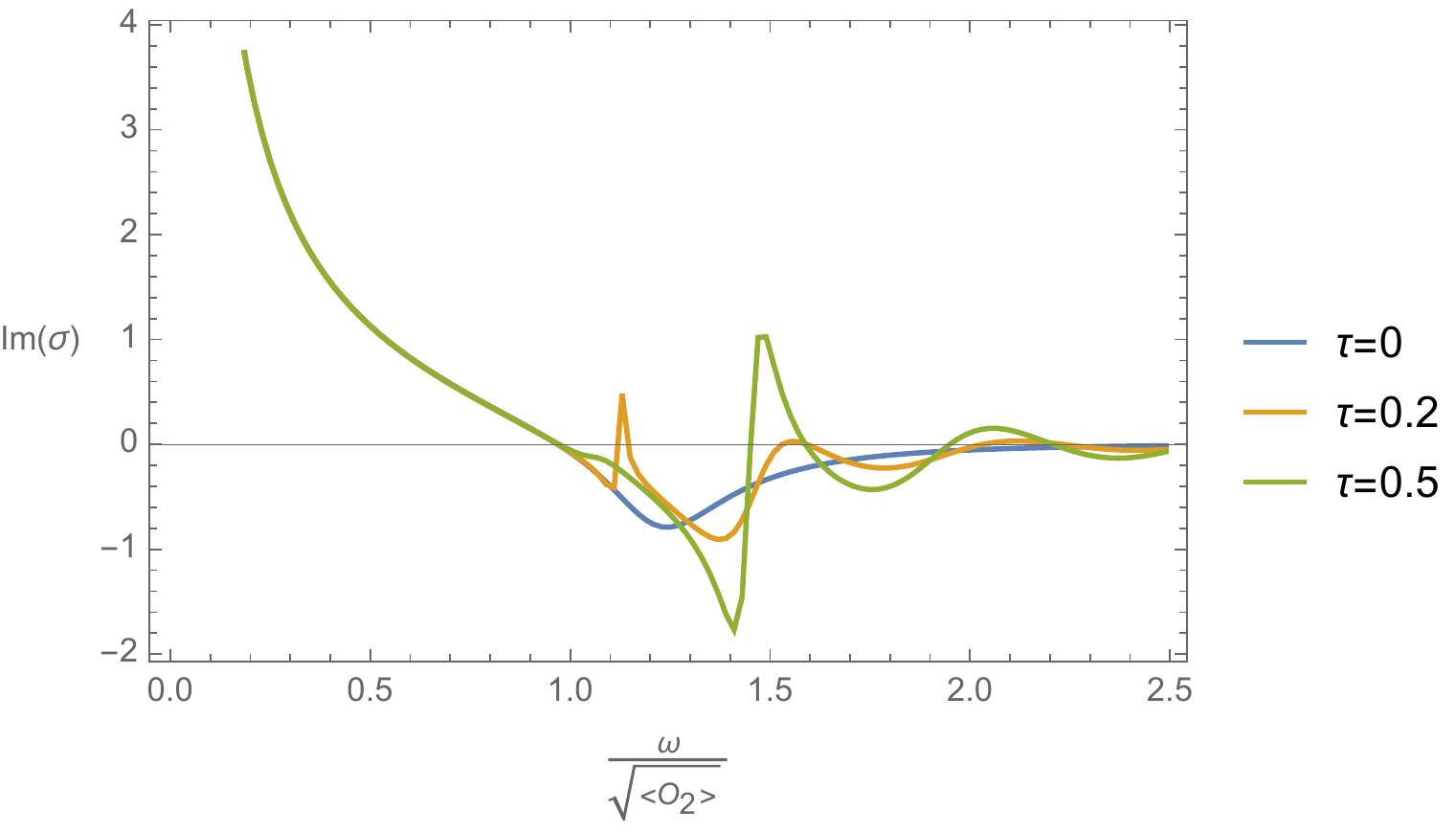}}
{\includegraphics[scale=0.5]{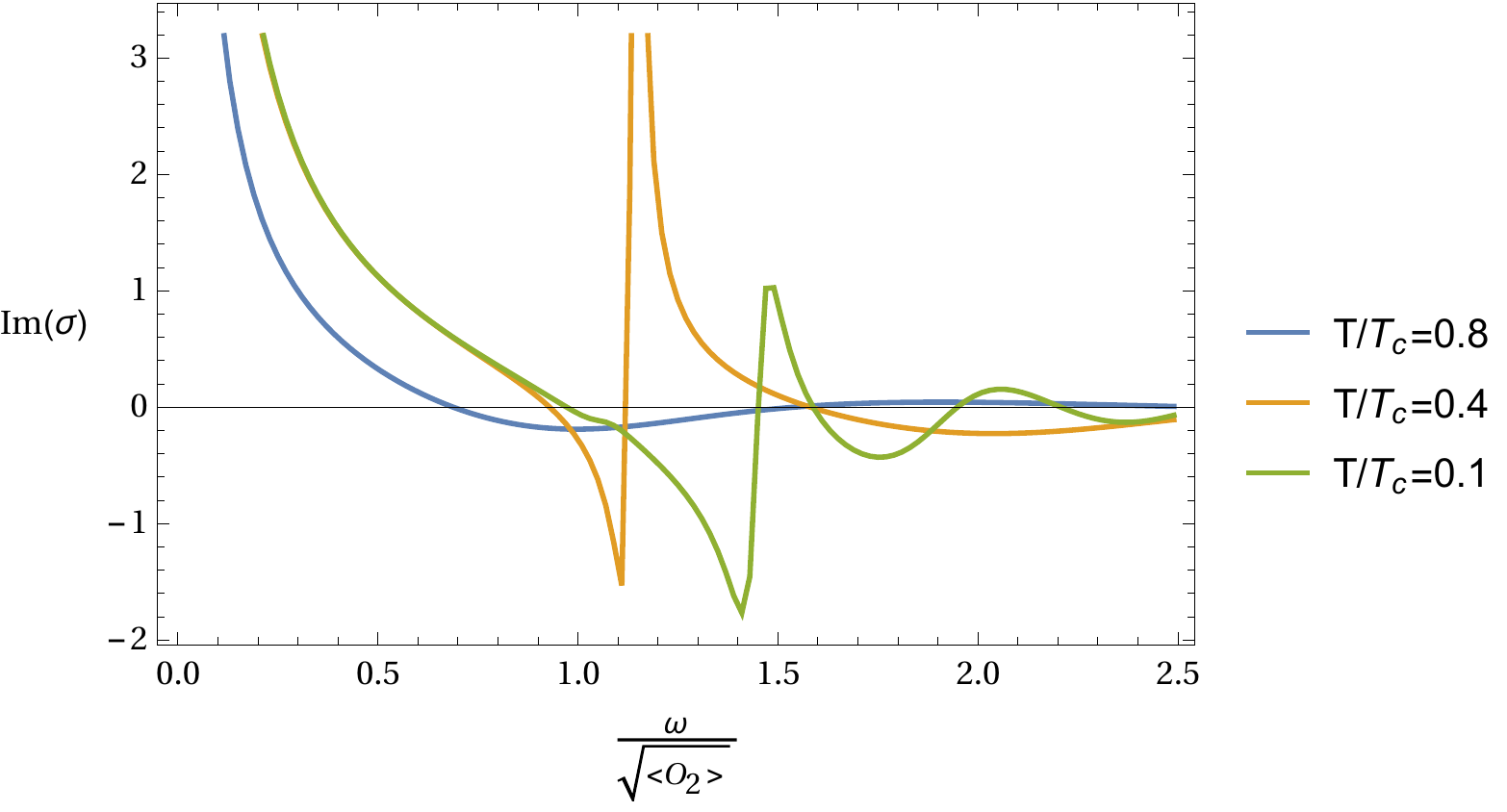}}
\caption{Real and imaginary part of conductivity. For the left diagrams $\beta=0.01$ and$\frac{T}{T_c}=0.1$. For the right diagrams $\beta=0.01$ and $\tau=0.5$.}
\label{fig4}
\end{figure}
\begin{figure}[h]
{\includegraphics[scale=0.5]{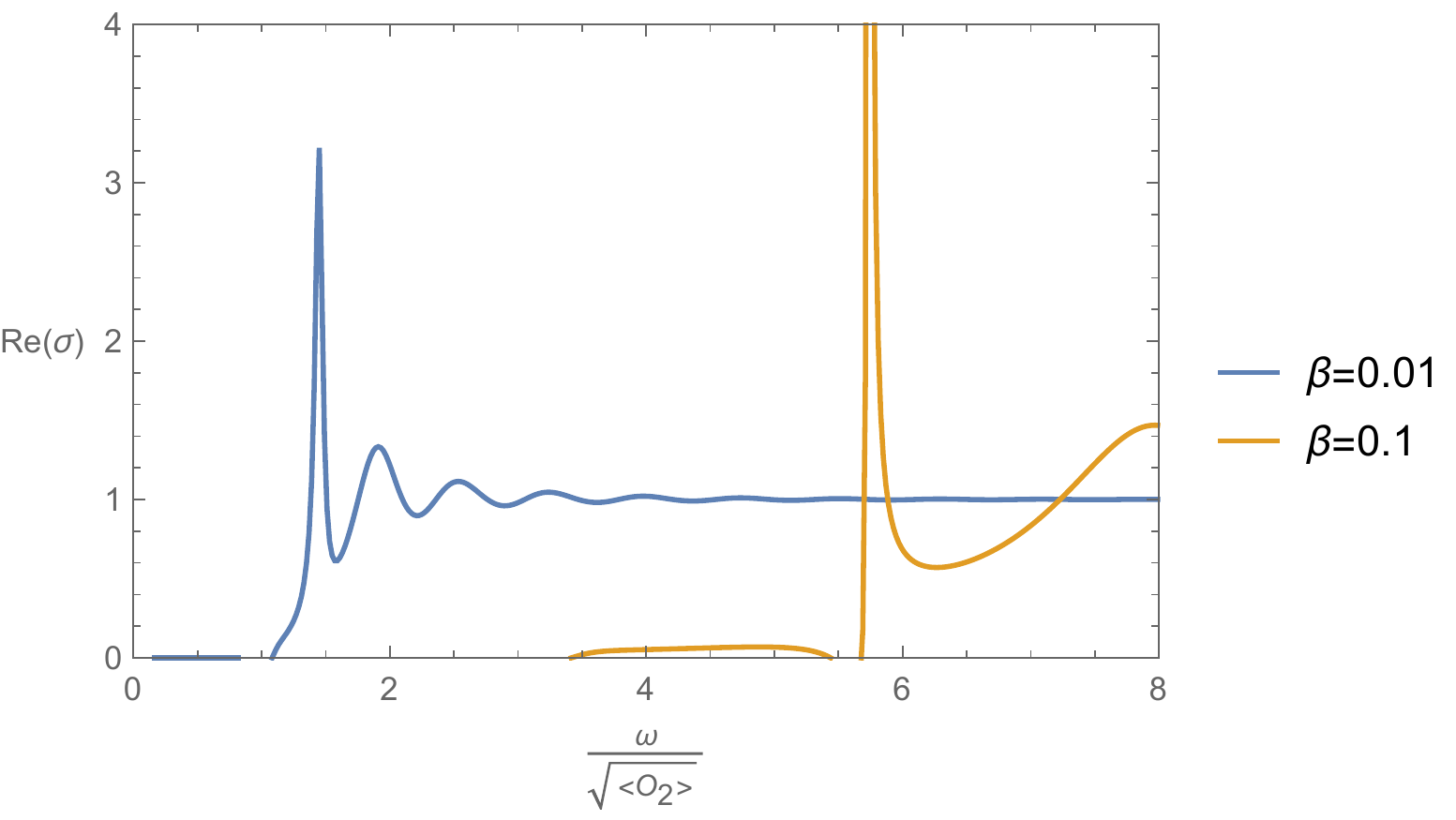}}
{\includegraphics[scale=0.5]{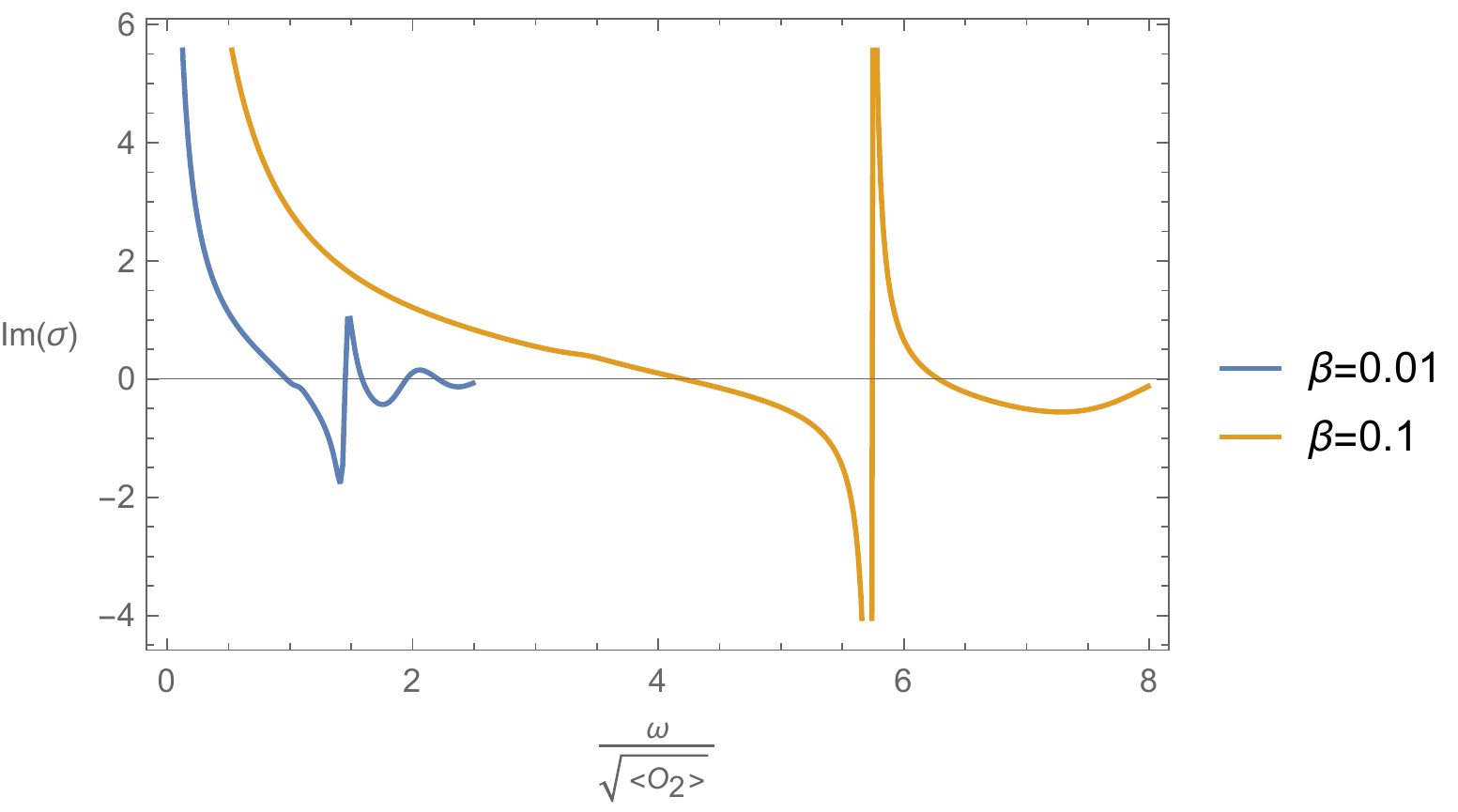}}
\caption{Real and imaginary part of conductivity for $\tau=0.5$ and $\frac{T}{T_c}=0.1$.} \label{fig5}
\end{figure}

We first fix the derivative  coupling $\beta=0.01$ and we vary the dissipative parameter $\tau$. We see in Fig.~\ref{fig4} (left panel) that when $\tau=0$ we do not have any peak. As the $\tau$ is increasing, we have a clear formation of peaks. In the right panel of Fig.~\ref{fig4} we see that as we go beyond the critical temperature $T_c$ in the supercondacting phase there is a clear formation of peaks. Finally in Fig.~\ref{fig5} we fix the dissipative parameter and we vary the derivative coupling. We see that larger values of $\beta$ give a clear peak at larger $\omega$.

This behaviour is interesting showing that the AC conductivity shows a non-trivial structure indicating a collective excitation of the charge carriers.
Similar behaviour was observed in  \cite{Baggioli:2015zoa} in which the translational symmetry is broken by massive graviton effects, and also  in holographic models that behave effectively as prototypes of Mott insulators  \cite{Baggioli:2016oju}. A study of competition of various phases was carried out in \cite{Kiritsis:2015hoa,Baggioli:2015dwa} using as control parameters the temperature and a doping-like parameter.  This coexistence of phases can be seen more clearly if you are away from the critical temperature where due to the proximity effect there is a leakage of Cooper pairs to the normal phase. There are two competing effects. The first one is the momentum dissipation which restricts the kinetic properties of the charge carries and in the same time the derivative coupling which acts as a doping parameter magnifying the effect of the momentum dissipation. To have a better understanding of these effects in the next sections we will study the conductivity in the fully backreacting theory.

\section{Spontaneous breaking of translational invariance near the criticality}\label{sec-criticality}

In this section we will discuss the critical temperature below which the black hole solution will develop hair and the way we break
 spontaneously  the translational invariance near the criticality. Our aim is to go beyond the probe limit and find a fully backreacted
 hairy black hole solution. In general the resulted hairy solution could have two
possible sources of instabilities, a negative mass for the scalar field and from the  propagating modes of the scalar field coming from its kinetic energy.
In the context of the $AdS/CFT$ correspondence the first one is known as the Breitenlohner-Freedman (BF) bound \cite{Breitenlohner:1982jf,Mezincescu:1984ev}.
However, this bound corresponds to  a conformally coupled scalar in
the background of a Schwarzschild AdS black hole, and arises in contexts in which the $AdS_4/CFT_3$ correspondence
is embedded into string theory and M theory. We should note that our Lagrangian resulted from the action
(\ref{EGBscalaraction}) is not clear how it arises from string or M theory. On the other hand even if the BF bound is satisfied
it does
not guarantee the nonlinear stability of hairy black holes under general boundary conditions
and potentials as it was discussed in \cite{Koutsoumbas:2009pa}.

For the other source of possible instabilities in  \cite{Kolyvaris:2011fk,Kolyvaris:2013zfa}  fully backreacted black hole solutions were found in the presence of the derivative coupling $\beta$. However, these solutions exists only in the case of a negative sign of the derivative coupling while if the coupling constant $\beta$ is positive, then the system of Einstein-Maxwell-Klein-Gordon equations is unstable and no solutions were found.
  In \cite{Koutsoumbas:2013boa} a very small window of positive $\beta$ was shown to be allowed. For positive derivative coupling the stability of the Galileon black holes was investigated but there is no any conclusive result. For example in \cite{Minamitsuji:2014hha,Huang:2018kqr} the black hole
quasinormal modes in a scalar-tensor theory with the scalar field coupled
to the Einstein tensor were calculated. In the following we will calculate the BF bound for the action (\ref{EGBscalaraction}) and in the next section we will investigate the effects of the derivative coupling $\beta$ of both signs.

We consider the following ansatz for the metric, Maxwell and scalar field respectively
\bea\label{eq:ansatz}
ds^2&=&-f(r)dt^2+\frac{dr^2}{f(r)}+e^{2V_1(r)}dx^2+e^{2V_2(r)}dy^2~,\nonumber\\
A&=&\phi(r)dt~,\nonumber\\
\Psi&=&e^{ikx}\varphi(r)~,
\eea
where the functions $f, V_1,V_2, \phi,\varphi$ are all to be determined. In the normal phase, we have $\Psi=0$ and the  usual Reissner-Nordstr\"om black hole is a solution to the field equations where
\begin{align}\label{eq:RNsolution}
&V_1(r)=V_2(r)=\log r~, \nonumber\\
&f(r)\,=r^2\left(1-\left(r_h^2+\frac{\mu^2}{4}\right)\frac{r_h}{r^3}+\frac{r_h^2\mu^2}{4r^4}\right)\,~, \nonumber\\
&\phi(r)\,=\mu\left(1-\frac{r_h}{r}\right)\,~.
\end{align}
The Hawking temperature is
\begin{equation}
T\,=\,\frac{3 \,r_h}{4\, \pi }-\frac{\mu ^2}{16 \,\pi\,r_h }~.
\end{equation}
The near horizon extremal solution is the usual $AdS_2\,\times\,\mathbb{R}^2$ where the extremal horizon is located at
$r_h\,=\,\frac{\mu}{2\sqrt{3}}~$ and the AdS$_2$ radius is $L_2^2\,=\,\frac{L^2}{6}~$.

From the Klein-Gordon equation \eqref{glgord} in the background of \eqref{eq:ansatz}, we see that the scalar $\Psi$ gets an effective momentum dependent  mass
\begin{equation}
m_{eff}^2(r)\,=\,m^2\,+\,k^2\,C^{xx}\,=\,m^2\,+k^2\left(\frac{1}{r^2}+\frac{\beta(2f'+r f'')}{2r^3}\,\right)~.
\end{equation}
Further considering the expression of \eqref{eq:ansatz},   we obtain that an instability in the near horizon region appears when
\begin{equation}
m_{eff}^2(r_h)L_2^2\,=\frac{L^2}{6}\left(m^2+\frac{12 (6 \beta +1) k^2}{\mu ^2}\right)<\,m_{BF}^2=-\frac{1}{4}(1+3\beta)~.
\end{equation}
Moreover, to have a well-defined theory, $\beta>-1/3$ is required to satisfy $m_{BF}^2<0$.  Thus, to fulfill the above conditions,  the range of $\beta$ is
\begin{equation}
-\frac{1}{3}<\beta<-\frac{1}{6}~,
\end{equation}
which obviously decreases the  effective mass.

\begin{figure}[h]
\includegraphics[scale=0.7]{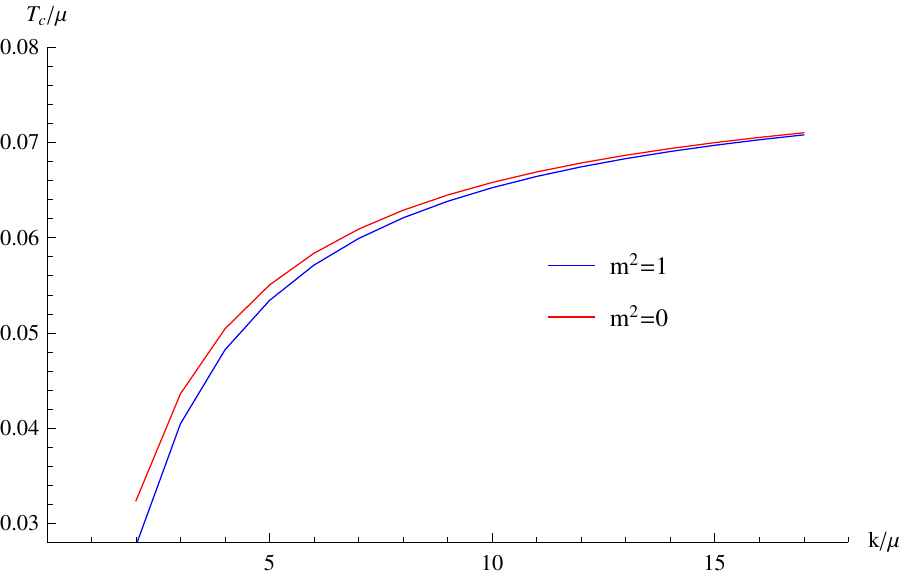}\hspace{0.5cm}
\caption{ The critical temperature v.s. the wavenumber.} \label{fig-Tc}
\end{figure}

In order to find the critical temperature, we have to solve  the Klein-Gordon equation \eqref{glgord} with the ansatz $\Psi\,=\,e^{i k x} \varphi(r)$
\begin{eqnarray}
\left(1+\beta\left(\frac{f}{r^2}+\frac{f'}{r}\right)\right)\varphi''+\left(\frac{2}{r}+\frac{f'}{f}+\beta\left(\frac{3f'}{r^2}+\frac{f'^2}{rf}+
\frac{f''}{r}\right)\right)\varphi'-\left(\frac{m^2}{f}+\frac{k^2}{r^2f}+\beta k^2\left(\frac{f'}{r^3f}+\frac{f''}{2r^2f}\right)\right)\varphi=0
\end{eqnarray}
in the background of the Reissner-Nordstr\"om black hole. We have to look for a normalizable solution of the field $\varphi$ where the source of $\varphi$ is zero but the vacuum expectation  value is not.
Since the boundary condition of $\varphi$ is the same as shown in \eqref{eq-boundayM},
 we will find solutions with $\varphi_-=0$ and $\varphi_+\neq 0$.
We then integrate the equation numerically shooting from the horizon and we search for a normalizable solution. In particular, we fixed the value of the derivative coupling to  $\beta=-1/4$, and find out the critical temperature $T_c(\beta)$ corresponding to different momentum. The results are shown in Fig.~\ref{fig-Tc}.  We see that  the lowest critical temperature is very small, but certainly happens at a non-zero value of momentum. As $k$ becomes larger, $T_c$ increases monotonously, which means that the maximum critical temperature may go to be  infinity. The critical temperature does not have the  bell-shaped behavior as it was expected  \cite{Donos:2011bh}.

 However, to find a finite critical temperature and produce a bell-shaped diagram, a method of regularization was proposed in \cite{Alsup:2013kda} where another term with higher derivatives of the scalar field coupled to Einstein tensor
\be  - \beta' | D_\mu G^{\mu\nu} D_\nu \Psi |^2 \  \label{int_Lag_den}\ee
was added into the action (\ref{EGBscalaraction}).
Then by solving the radial Klein-Gordon equation by setting $\Psi(r,x)=\varphi(r) \cos(kx)$, the authors  concluded that  when $\frac{\beta}{\mu^2}=\frac{\beta'}{\mu^4} =0$,  the maximum critical temperature $T_c$ is found  at $k=0$, which shows that the homogeneous solution is dominant. Turning on the higher-derivative  interaction terms, the $T_c$ of the system  depends on the coupling constants $\beta$, $\beta'$, and the homogeneous solution no longer dominates. Then it was found that for $\beta$ large enough while $\beta' =0$, the asymptotic ($k^2/\mu^2 \to \infty$) transition temperature will be higher than that of the homogeneous solution. In this case, the transition temperature monotonically increases as we increase the wavenumber $k$ having the same behaviour we found in Fig.~\ref{fig-Tc}. As they switched on $\beta' >0$, the transition temperature  attains a maximum value at a finite $k$ as it was shown in the Fig.~$2$ of \cite{Alsup:2013kda}. Thus the second higher-derivative coupling acts as a UV cutoff on the wavenumber,

The physical explanation of this behaviour is that the presence of the first derivative coupling to Einstein tensor $\beta$, is the encoding of the electric field's back reaction near the horizon and it is the  cause of spontaneous generation of spatial modulation, while the presence of the second higher derivative coupling $\beta'$ can be understood as stabilizing the inhomogeneous modes introduced by the
first derivative coupling.

In the next section we will find numerically a fully backreacted hairy black hole solution of the charged scalar field coupled to Einstein tensor with the coupling $\beta$, assuming that near the
critical temperature $T \approx T_c$ the effects of the cutoff are negligible  and set $\beta'=0$.
 Then, trusting  our linearization below the critical temperature because it will not rely on a gradient expansion but on an order parameter proportional to $\left(T-T_c\right)^{1/2}$, we will calculate the conductivity.

\section{Hairy black hole solutions beyond the probe limit}\label{sec-backreaction}

\subsection{The hairy black hole  with full backreaction}\label{sec-backreaction1}

To find the  black hole solution with full backreaction, we consider the ansatz of the fields as
\begin{align}\label{eq-Metric}
ds^2 &= \frac{1}{z^2} \left( -U(z)dt^2 + \frac{1}{U(z)} dz^2 + V_1(z)dx^2 + V_2(z)dy^2 \right)~,\nonumber \\
A &= A_t(z)dt~, \nonumber\\
\Phi &= e^{ikx}\varphi(z)~,
\end{align}
then the horizon is located at $z=1$ and the asymptotical boundary is at $z\to 0$ . With this ansatz,  the coupled Einstein-Maxwell-scalar field equations (\ref{einst}), (\ref{glgord}) and (\ref{max}) become a set of coupled differential equations which are given in the Appendix A.

To solve  numerically the highly coupled system, we need to analyze the asymptotical solution of the scalar field $\varphi$ in the Klein-Gordon equation (\ref{glgord}) or its equivalent equation \eqref{eq-boundayM} with the coordinate transformation $z=r_h/r$. In order to simplify the asymptotical behavior of the scalar field, we will choose
\begin{equation} \label{mass}
 m^2=-2(1+3 \beta) \, ,
\end{equation}
 so that $\varphi |_{z \rightarrow 0}  = \varphi_- z+\varphi_+z^2~.$
Furthermore, as pointed out in \cite{Horowitz:2012ky, Andrade:2017jmt},  it is convenient to define
\begin{eqnarray}
\nonumber U(z) &=&  (1-z) u(z) \, , \\
\nonumber A_t(z) &=& (1-z) a(z) \, , \\
\varphi (z) &=& z \psi (z) \, ,
\end{eqnarray}
so, we will obtain numerically the functions $u(z)$, $\psi (z)$, $a(z)$, $V_1(z)$ and $V_2(z)$ with appropriate boundary conditions.
\begin{figure}[h]
\begin{center}
\includegraphics[width=50mm]{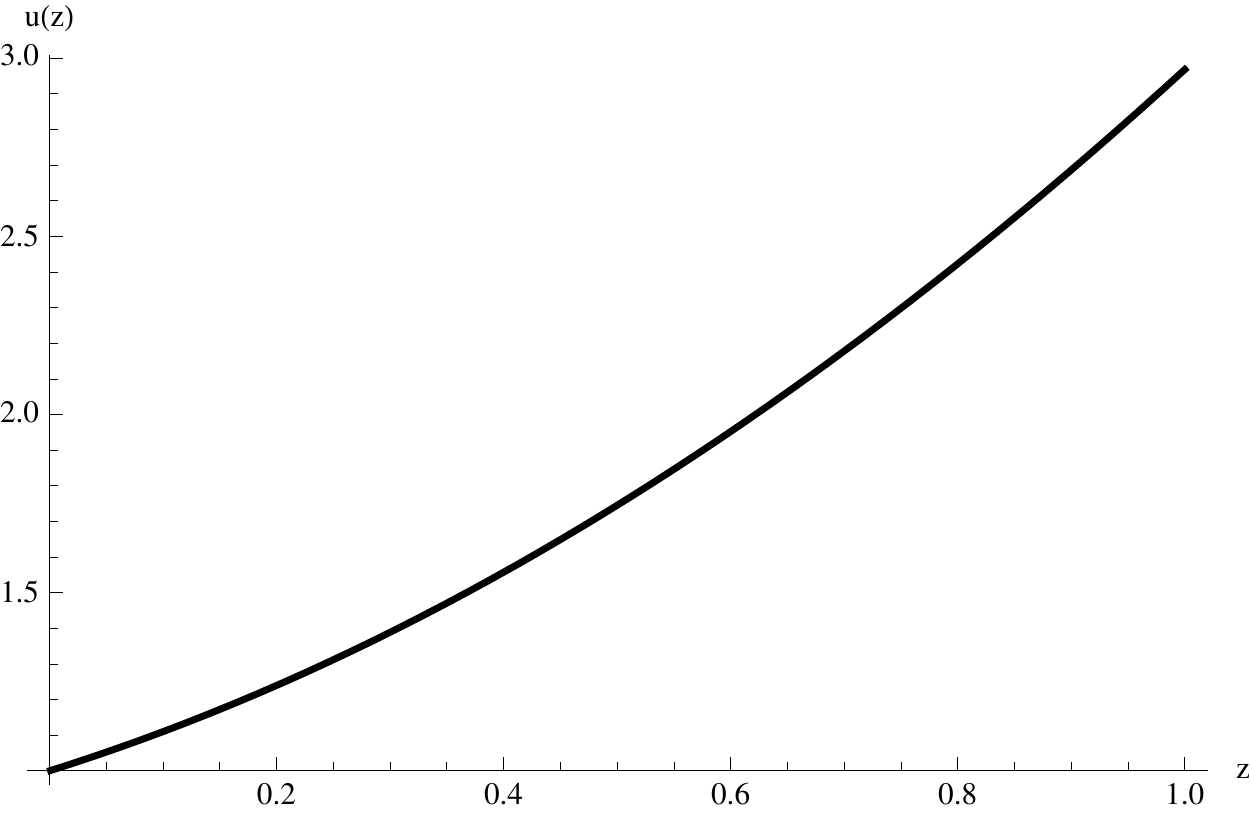}\hspace{0.8cm}
\includegraphics[width=50mm]{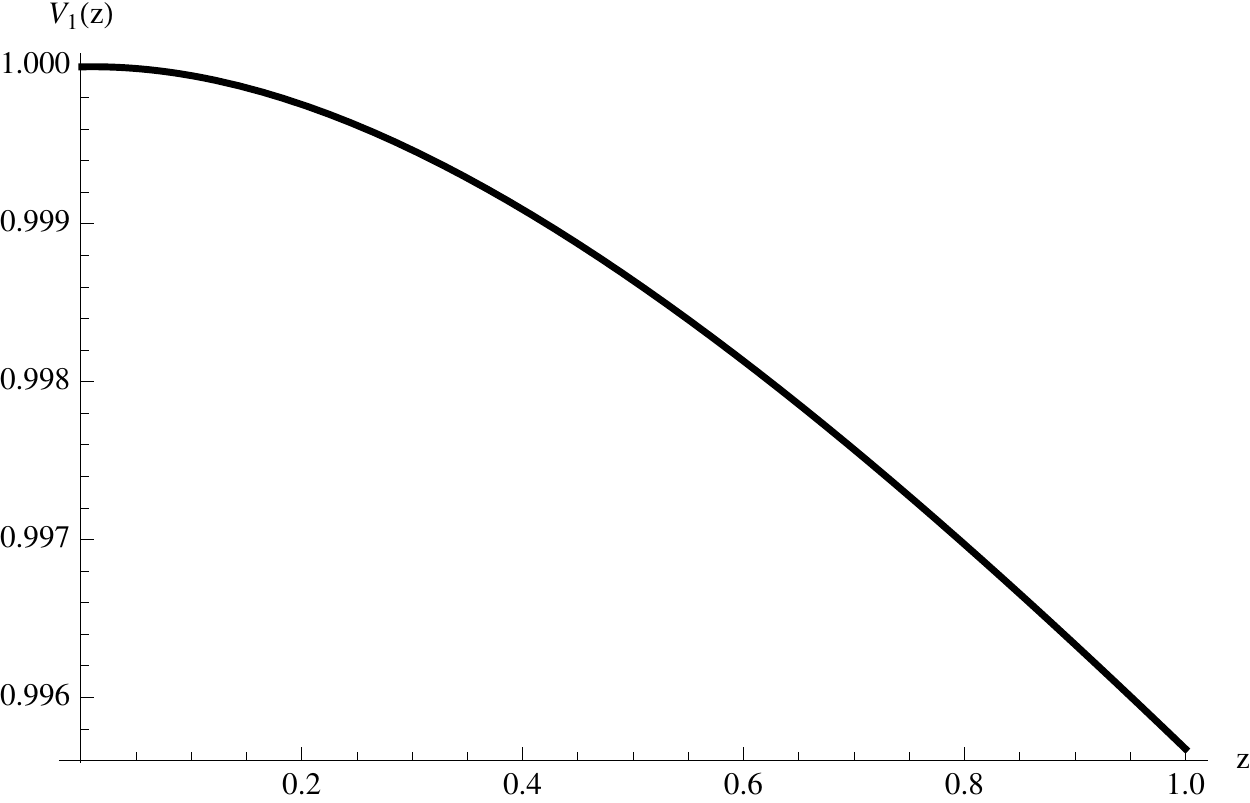}\hspace{0.8cm}
\includegraphics[width=50mm]{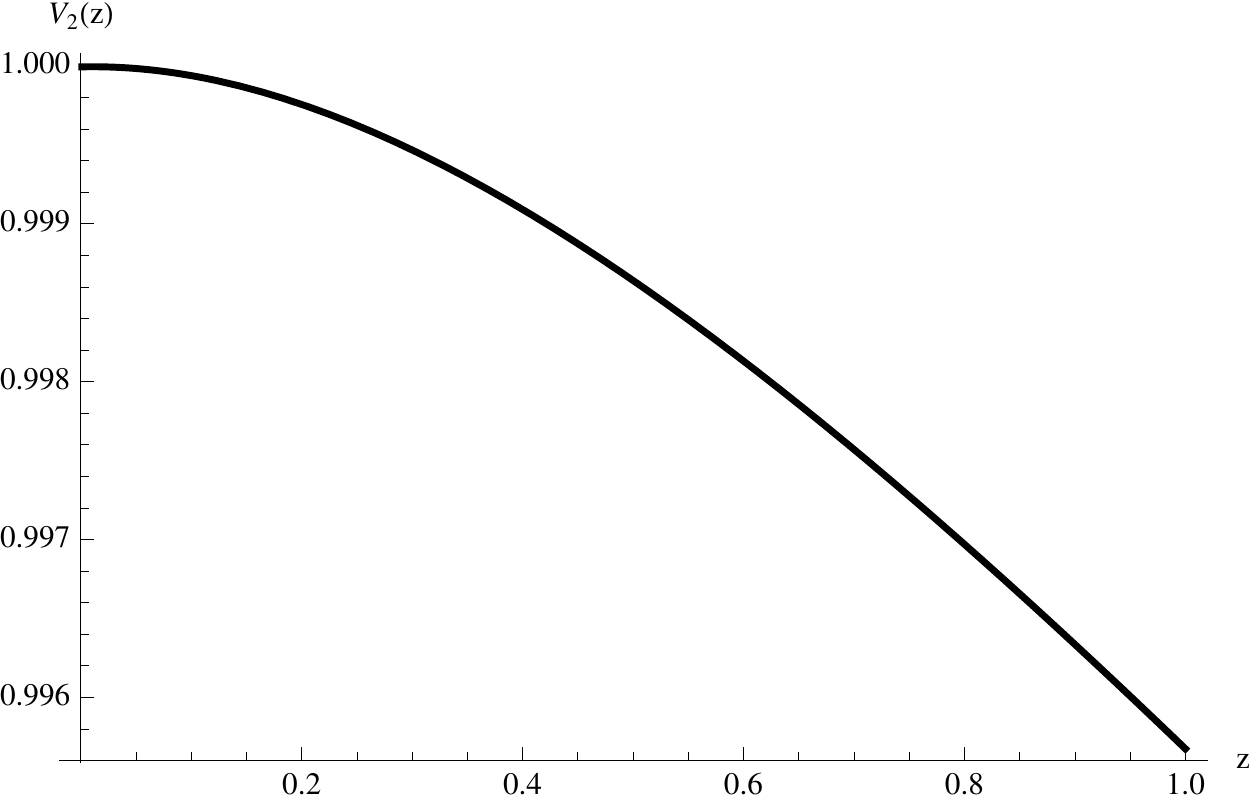}
\end{center}
\caption{The behavior of the fields $u(z)$,  $V_1(z)$ and $V_2(z)$ as a function of $z$ for $\beta=0.1$, $\lambda / \mu =0.3$, $k / \mu = 0$ and $T / \mu = 1$.}
\label{fig-profile1}
\end{figure}

\begin{figure}[h]
\begin{center}
\includegraphics[width=50mm]{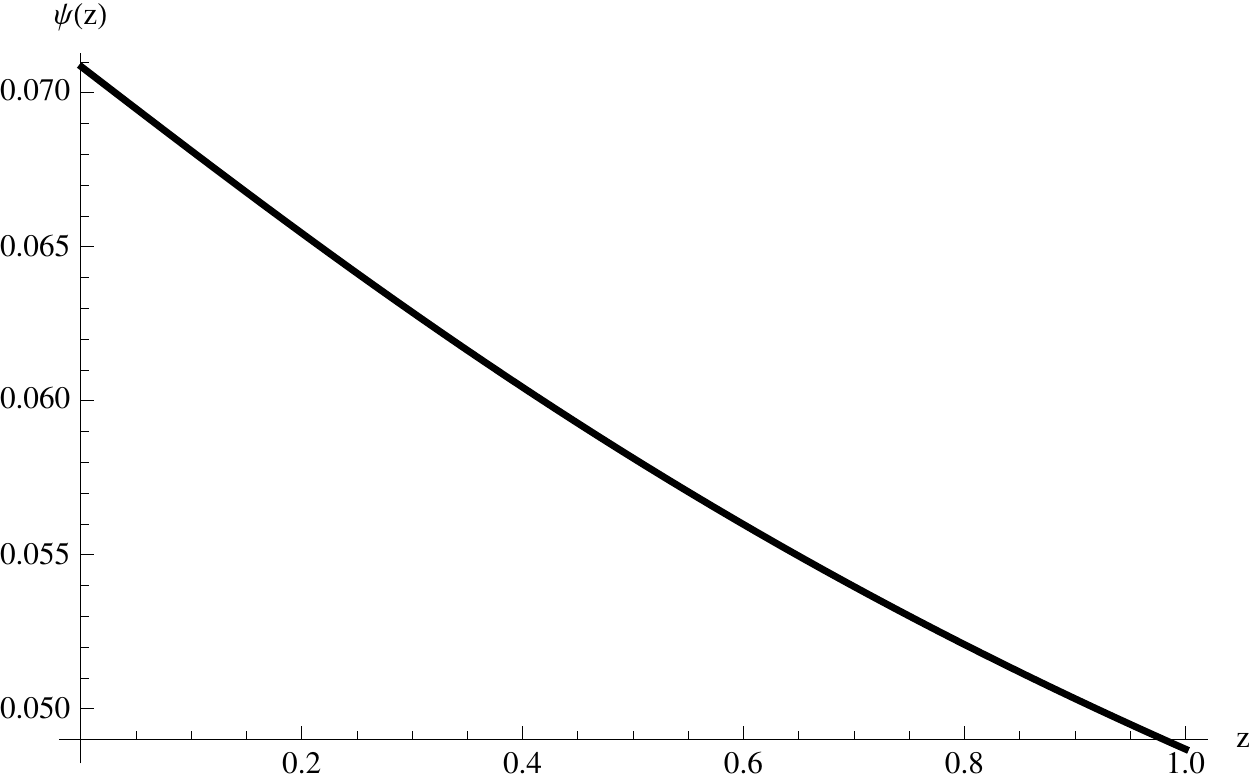}\hspace{0.5cm}
\includegraphics[width=50mm]{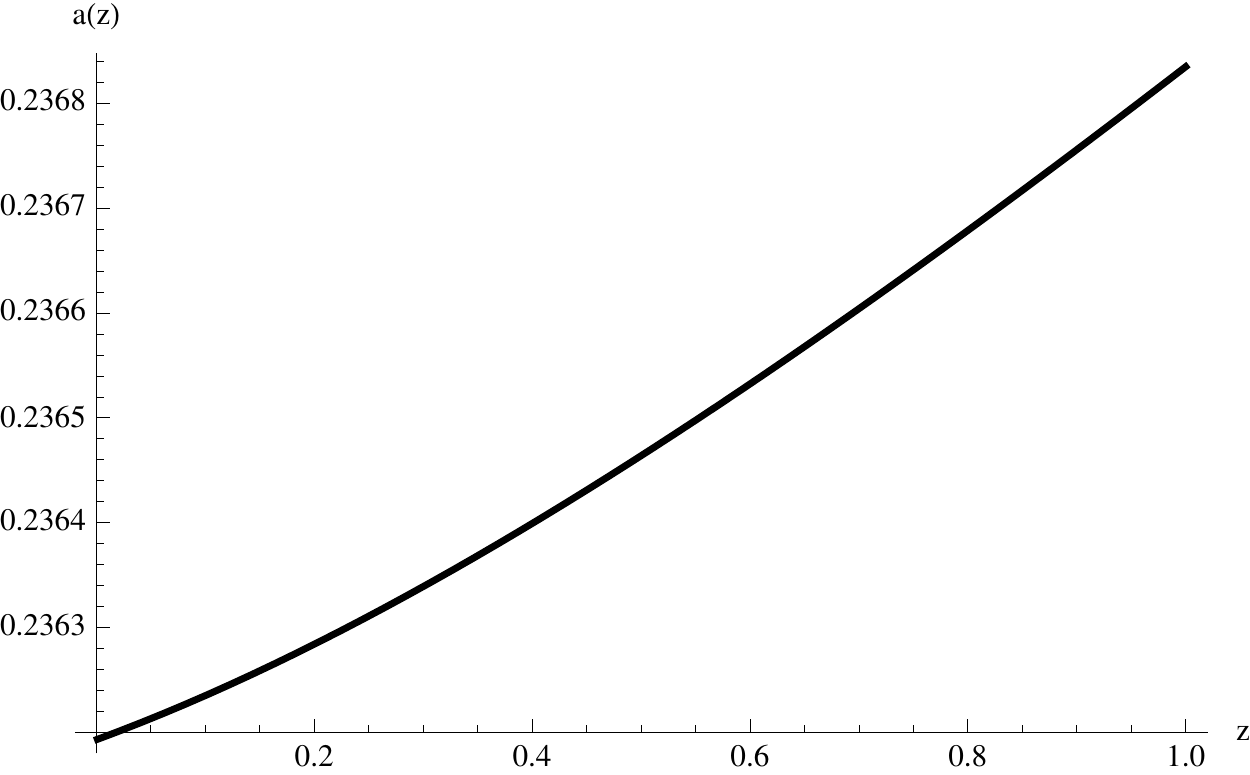}\hspace{0.5cm}
\end{center}
\caption{The behavior of the fields  $\psi (z)$, $a(z)$ as a function of $z$ for $\beta=0.1$, $\lambda / \mu =0.3$, $k / \mu = 0$ and $T / \mu = 1$.}
\label{fig-profile2}
\end{figure}

\begin{figure}[h]
\begin{center}
\includegraphics[width=50mm]{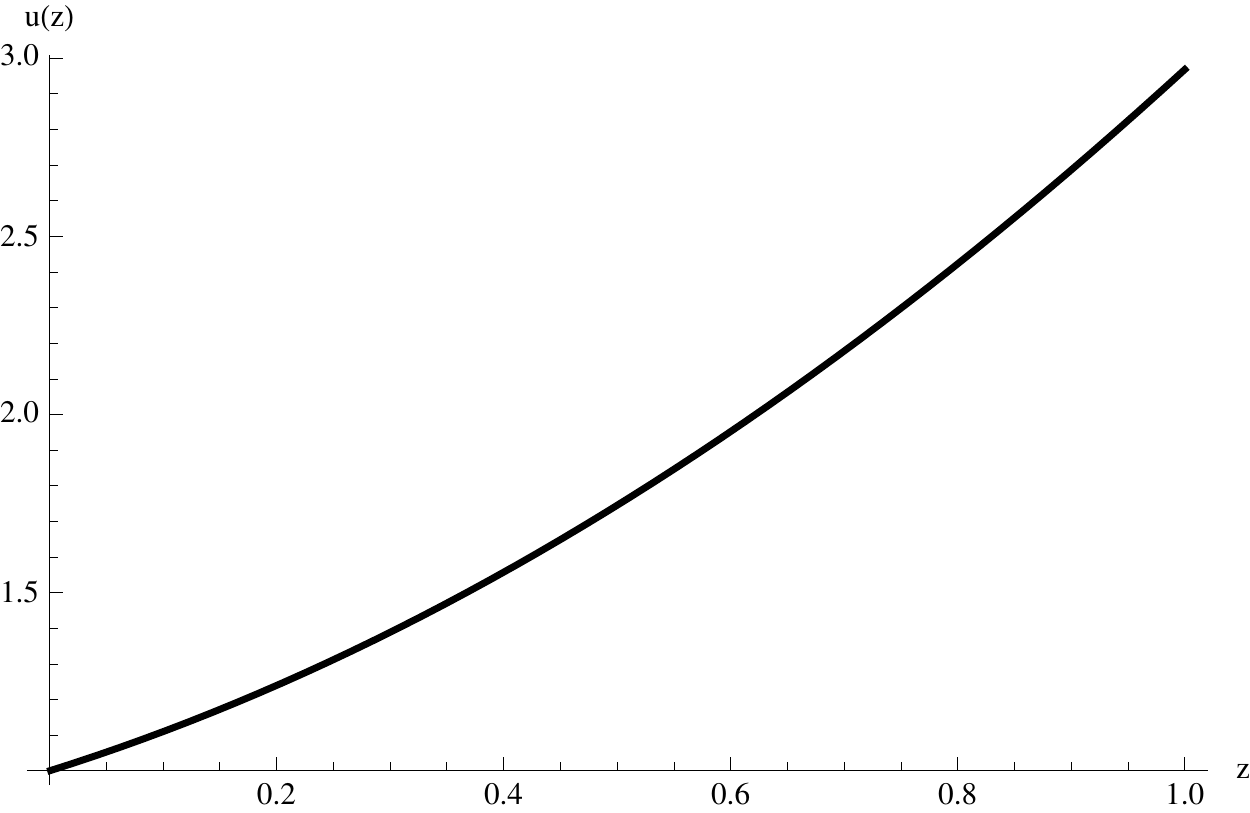}\hspace{0.8cm}
\includegraphics[width=50mm]{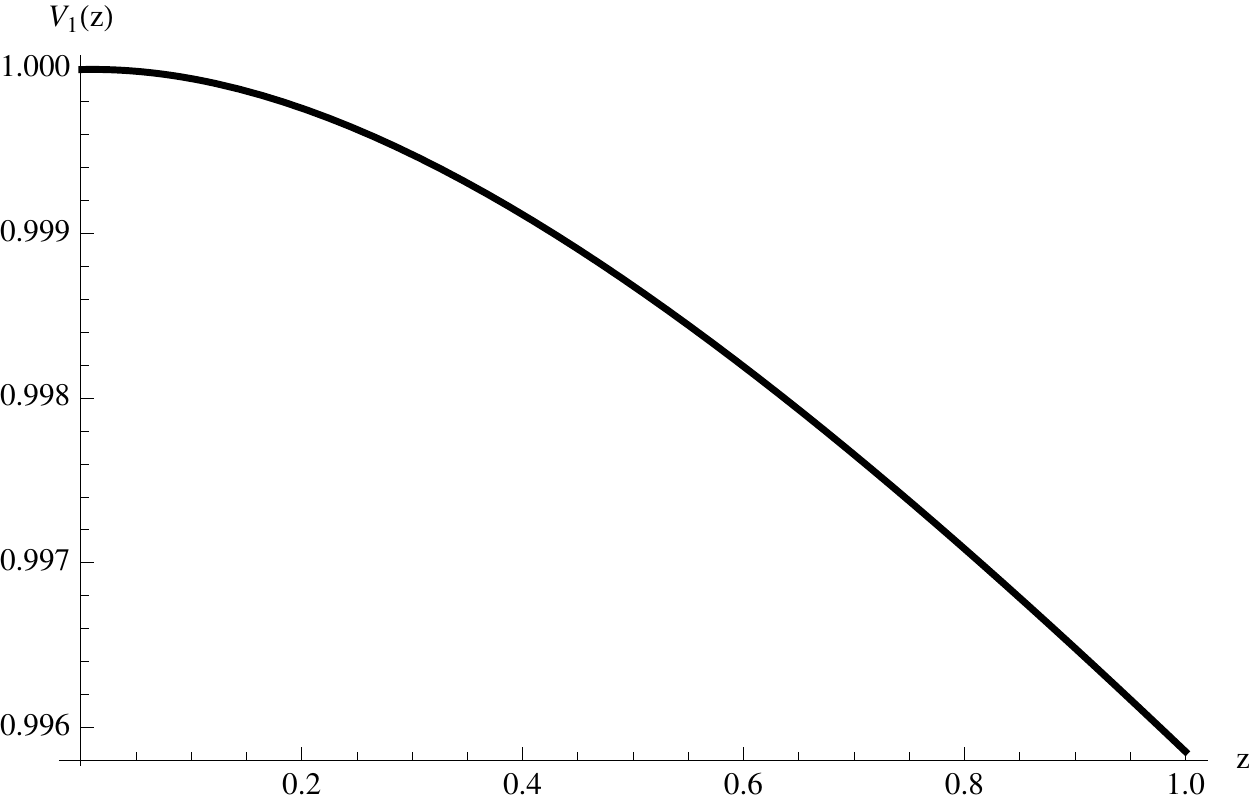}\hspace{0.8cm}
\includegraphics[width=50mm]{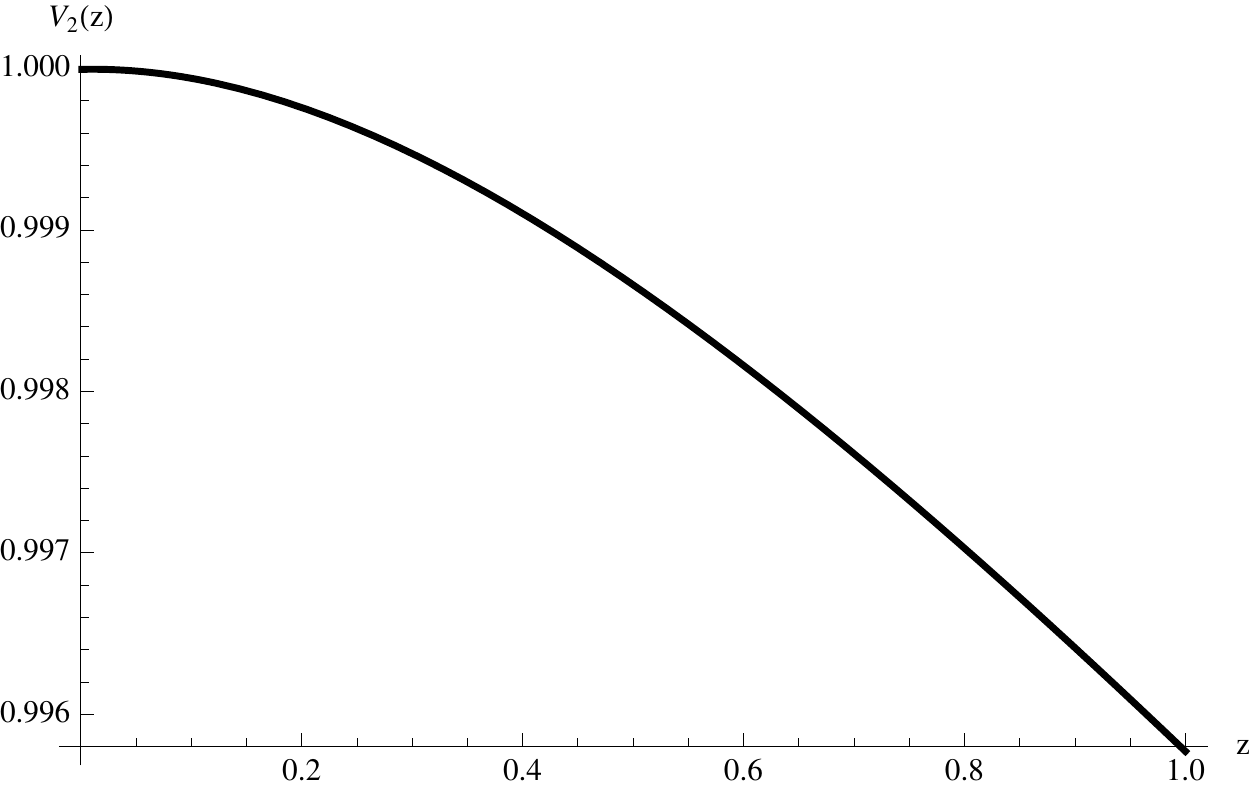}
\end{center}
\caption{The behavior of the fields $u(z)$,  $V_1(z)$ and $V_2(z)$ as a function of $z$ for $\beta=0.1$, $\lambda / \mu =0.3$, $k / \mu = 1$ and $T / \mu = 1$.}
\label{fig-profile3}
\end{figure}

\begin{figure}[h]
\begin{center}
\includegraphics[width=50mm]{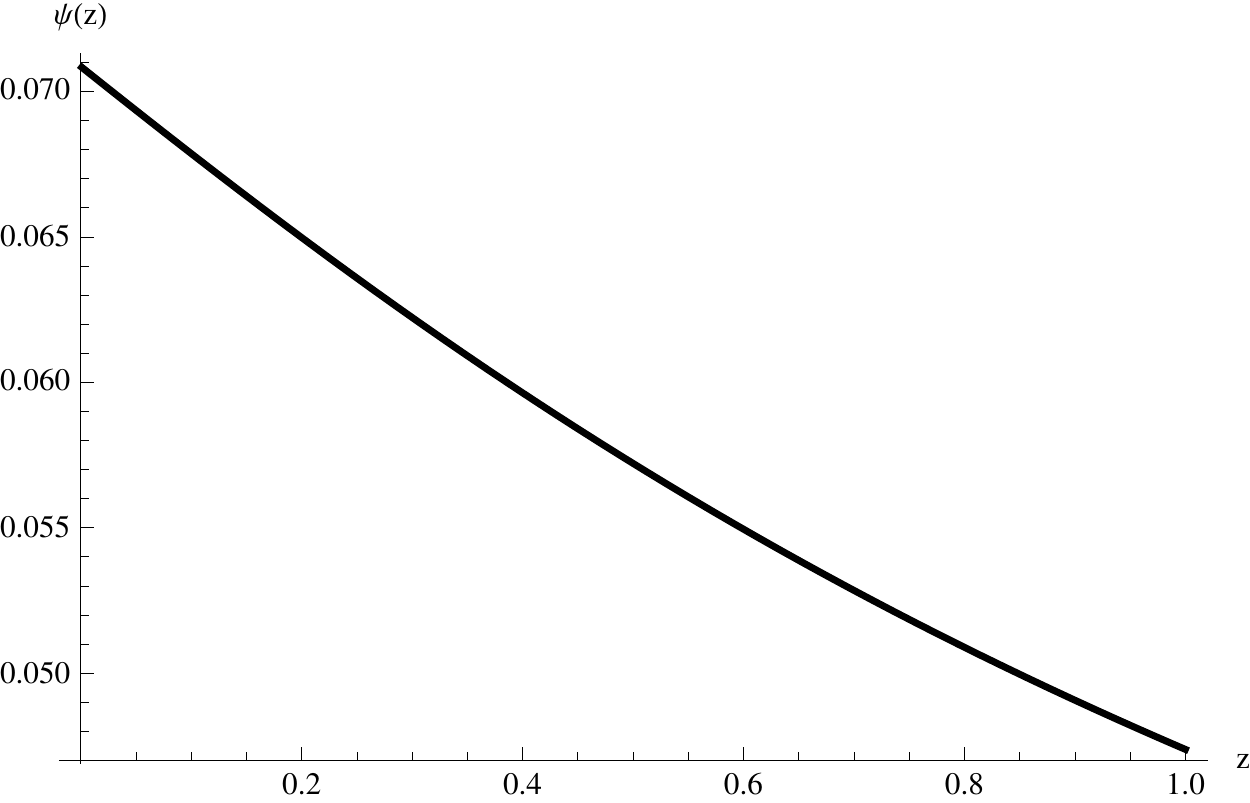}\hspace{0.5cm}
\includegraphics[width=50mm]{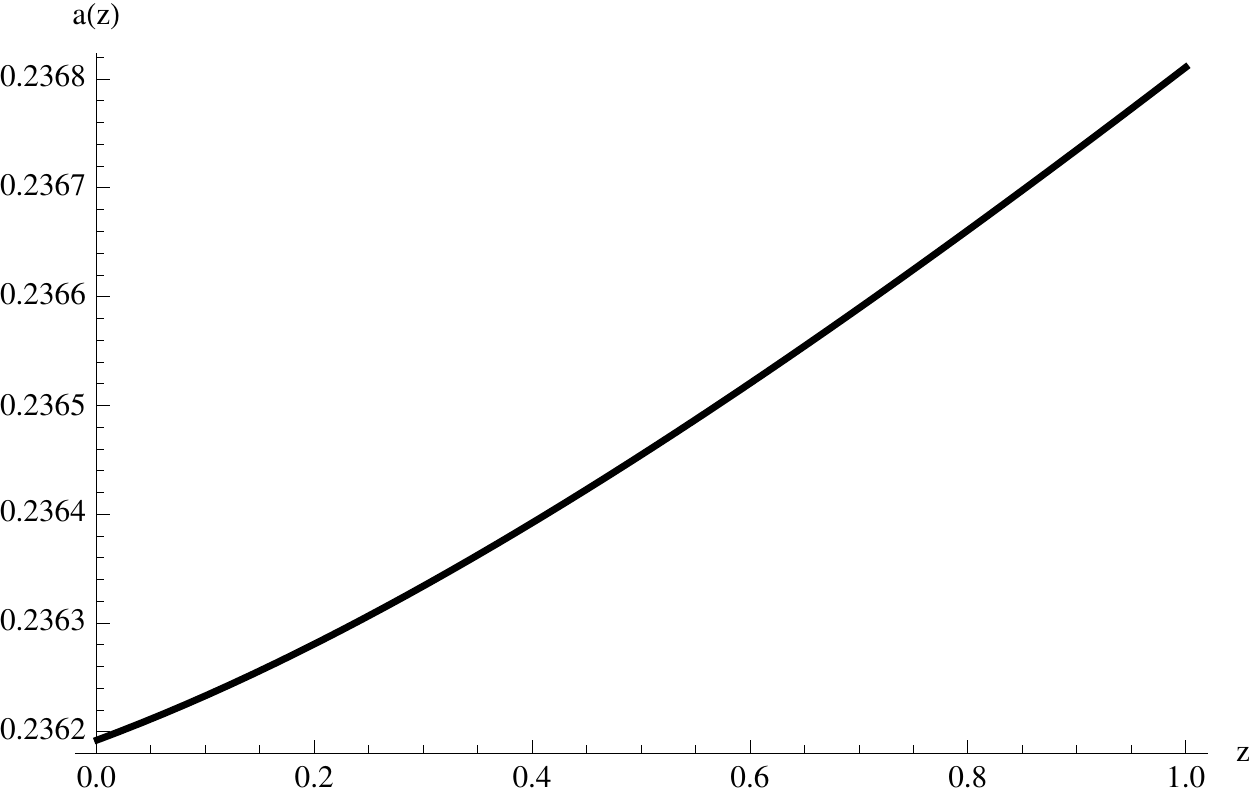}\hspace{0.5cm}
\end{center}
\caption{The behavior of the fields  $\psi (z)$, $a(z)$ as a function of $z$ for $\beta=0.1$, $\lambda / \mu =0.3$, $k / \mu = 1$ and $T / \mu = 1$.}
\label{fig-profile4}
\end{figure}

The boundary conditions at radial infinity ($z \rightarrow 0$) read
\begin{equation}
u(z)=-\frac{\Lambda}{3}+\mathcal{O}(z)\, , \,\, \psi(z)=\lambda + \mathcal{O}(z)\, , \,\,  a(z)= \mu-\rho z  + \mathcal{O}(z^2) \, , \,\, V_1(z)= 1+ \mathcal{O}(z) \, , \,\, V_2(z)= 1+  \mathcal{O} (z)~,
\end{equation}
and we take $\Lambda=-3$. We impose regularity condition for all the fields  at the horizon. The  boundary conditions for the numerics we imposed, are similar  to the boundary conditions which were discussed in details in \cite{Donos:2013eha}. In our case
 we have an additional $\beta$ coupling term, but this will not modify the boundary conditions after we propose the relation \eqref{mass}. Thus,
for fixed $m^2$, our  theory is also  specified by three dimensionless parameters $T/ \mu$, $\lambda/ \mu $ and $k/ \mu$ which are similar to the ones in \cite{Donos:2013eha}.

Profiles of the fields are shown in Fig.~\ref{fig-profile1} - Fig.~\ref{fig-profile4} for a choice of parameters. These figures show that we have found hairy black hole solutions with $k = 0$ (Fig.~\ref{fig-profile1}, Fig.~\ref{fig-profile2}) or $k\neq 0$ (Fig.~\ref{fig-profile3}, Fig.~\ref{fig-profile4}) having the translational symmetry preserved or broken. We observe that in both cases the found hairy black hole solutions have the same behavior and the scalar field is regular on the horizon.

\subsection{Thermodynamics}
Now we study the thermodynamics of the black holes we numerically obtained. We shall use the Euclidean formalism \cite{Gibbons:1976ue}. This method has been used in the study of the thermodynamics of black holes in Horndeski gravity in Refs. \cite{Anabalon:2013oea, Bravo-Gaete:2014haa, Feng:2015oea, Sebastiani:2018hak}. The Euclidean and Lorentzian action are related by $I_{E}=-iI$, and the Euclidean time is $\tau=it$.
 The Euclidean continuation of the metric \eqref{eq-Metric} is
 \begin{equation}\label{solution}
ds^2=\frac{1}{z^2} \left( N^2(z) U(z) d\tau^2+\frac{1}{U(z)} dz^2 +V_1(z) dx^2+V_2(z) dy^2 \right) \,,
\end{equation}
and the electric potential to be considered is
\begin{equation}
A=A_{\tau}(z) d\tau \,,
\end{equation}
 also, we consider the scalar field is given by $\Phi=e^{ikx} \varphi(z)$ and we hold $k$ fixed in the following. Requiring the absence of conical singularity at the horizon in the Euclidean solution (\ref{solution}), the Euclidean time must be periodic, with period $\hat{\beta}=1/T$; therefore, the Hawking temperature is given by $T=-U'(1)/4\pi$. The Euclidean action evaluated for the metric (\ref{solution}), with $8 \pi G=1$, takes the Hamiltonian form
\begin{equation} \label{reduced}
I_{E}= -\hat{\beta} \sigma \int_{z=1}^{z=0}  dz \left( N \mathcal{H} +A_{\tau} \mathcal{G} \right) + B_{surf} \, ,
\end{equation}
where we have performed some integrations by parts. $\sigma$ is the area of the spatial 2-section, $\mathcal{G}=\partial_z \pi^z$, where $\pi^z=\frac{\sqrt{V_1 V_2}}{N}A_{\tau}^{\prime}$ is the conjugate momenta of the electromagnetic field, $\mathcal{H}$ is the reduced Hamiltonian, which is given by
\begin{eqnarray}
\notag \mathcal{H} &=& \sqrt{V_1 V_2} \Bigg( \frac{\beta  U \varphi ^2 k^2}{z^2 V_1}+\frac{\varphi ^2 k^2}{z^2 V_1}+\frac{3 \beta  U \varphi ^2 V_1'^2 k^2}{4 V_1^3}+\frac{2 \beta  U \varphi '^2 k^2}{V_1}+\frac{\beta  \varphi ^2 U' V_2' k^2}{4 V_1 V_2}+\frac{4 \beta  U \varphi  \varphi ' k^2}{z V_1}+\frac{\beta  \varphi  U' \varphi ' k^2}{V_1}+\frac{\beta  U \varphi  V_2' \varphi ' k^2}{V_1 V_2}+  \\
\notag  &&\frac{\beta  U \varphi ^2 V_2'' k^2}{2 V_1 V_2} + \frac{2 \beta  U \varphi  \varphi '' k^2}{V_1}- \frac{2 \beta  U \varphi  V_1' \varphi ' k^2}{V_1^2}-\frac{\beta  U \varphi ^2 V_1'' k^2}{2 V_1^2}-\frac{\beta  \varphi ^2 U' V_1' k^2}{4 V_1^2}-\frac{\beta  U \varphi ^2 V_1' k^2}{z V_1^2}- \frac{\beta  U \varphi ^2 V_1' V_2' k^2}{4 V_1^2 V_2}- \\
\notag && \frac{\beta  U \varphi ^2 V_2'^2 k^2}{4 V_1 V_2^2}+\frac{m^2 \varphi ^2}{z^4}+\frac{\beta  U^2 \varphi '^2}{z^2}+\frac{\beta  U^2 V_1'^2 \varphi '^2}{4 V_1^2}+ \frac{\beta  U^2 V_2'^2 \varphi '^2}{4 V_2^2}+\frac{U \varphi '^2}{z^2}+\frac{3 \beta  U U' \varphi '^2}{z}+\frac{\Lambda }{z^4}+\frac{3 U}{z^4}+\frac{U' V_1'}{4 z^2 V_1}+\frac{U' V_2'}{4 z^2 V_2}+ \\
\notag && \frac{U V_1' V_2'}{4 z^2 V_1 V_2}+\frac{U V_1''}{2 z^2 V_1}+\frac{U V_2''}{2 z^2 V_2}+\frac{4 \beta  U^2 \varphi ' \varphi ''}{z}-\frac{U'}{z^3}-\frac{\beta  U^2 V_1' \varphi ' \varphi ''}{V_1}- \frac{\beta  U^2 \varphi '^2 V_1''}{2 V_1}-\frac{3 \beta  U U' V_1' \varphi '^2}{4 V_1}-\frac{U V_1'}{z^3 V_1}-\frac{U V_1'^2}{4 z^2 V_1^2}-\\
  && \frac{\beta  U^2 V_2' \varphi ' \varphi ''}{V_2}-  \frac{\beta  U^2 \varphi '^2 V_2''}{2 V_2}-\frac{3 \beta  U U' V_2' \varphi '^2}{4 V_2}-\frac{U V_2'}{z^3 V_2}-\frac{\beta  U^2 V_1' V_2' \varphi '^2}{4 V_1 V_2}-\frac{U V_2'^2}{4 z^2 V_2^2} \Bigg)  + \frac{1}{2 \sqrt{V_1 V_2}} (\pi^z)^2 \, ,
\end{eqnarray}
and $B_{surf}$ is a surface term and is determined by considering that the Euclidean action has an extremum \cite{Regge:1974zd}. The equations of motion are determined varying the action with respect to the fields $N, U, V_1, V_2, \varphi$, $A_{\tau}$ and $\pi^z$. $N$ and $A_{\tau}$ are Lagrange multipliers in (\ref{reduced}), and variyng respect to them yields the Hamiltonian constraint $\mathcal{H}=0$, which is equivalent to the $t-t$ component of the gravitational equations, and the Gauss law $\partial_z \pi^z=0$, respectively; therefore, the value of the action on the classical solutions is just given by the surface term. It can be show that the field equations obtained by varying the Euclidean action with respect to the fields $N, U, V_1, V_2, \varphi$, $A_{\tau}$ and $\pi^z$ are equivalent to the original ones, and the metric function $N$ can be set to $N=1$. In order to have a well defined variational problem, the surface term $B_{surf}$ must cancel the surface terms that arise from the variation of the bulk term in the action (\ref{reduced}), and it is explicitly given by
\begin{eqnarray}
\notag \delta B_{surf} &=& -\frac{\hat{\beta} \sigma}{4 z^3 (V_1 V_2)^{5/2}}\Bigg[V_2 \delta \varphi ' \Big(-8 k^2 \beta  U V_1^2 V_2^2 \varphi  z^3+4 \beta  U^2 V_1^2 V_2^2 V_1' \varphi ' z^3+4 \beta  U^2 V_1^3 V_2 V_2' \varphi ' z^3-16 \beta  U^2 V_1^3 V_2^2 \varphi ' z^2 \Big)+ \\
\notag && V_2 \delta V_1' \Big(2 k^2 \beta  U V_1 V_2^2 \varphi ^2 z^3+2 \beta  U^2 V_1^2 V_2^2 \varphi '^2 z^3-2 U V_1^2 V_2^2 z \Big)+V_2 \delta V_1 \Big(-\beta  U V_1^2 V_2^2 U' \varphi '^2 z^3-  \beta  U^2 V_1 V_2^2 V_1' \varphi '^2 z^3   \\
\notag &&-k^2 \beta  V_1 V_2^2 \varphi ^2 U' z^3-3 k^2 \beta  U V_2^2 \varphi ^2 V_1' z^3+V_1^2 V_2^2 U' z+U V_1 V_2^2 \left(4 k^2 \beta  \varphi  \varphi ' z^2+4 k^2 \beta  \varphi ^2 z+V_1'\right) z -2z^3 V_1^2 V_2^2 A_{\tau} A_{\tau}'\Big)+ \\
\notag && V_2 \delta \varphi  \Big(4 k^2 \beta  V_1^2 V_2^2 \varphi  U' z^3+4 k^2 \beta  U V_1 V_2^2 \varphi  V_1' z^3-2 \beta  U V_1^3 V_2 U' V_2' \varphi ' z^3-2 \beta  U^2 V_1^2 V_2 V_1' V_2' \varphi ' z^3+8 \beta  U^2 V_1^2 V_2^2 V_1' \varphi ' z^2 \\
\notag && +8 \beta  U^2 V_1^3 V_2 V_2' \varphi ' z^2-2 \beta  U V_1^2 V_2^2 (8 \varphi  k^2+z (4 k^2+U' V_1') \varphi ') z^2-8 U V_1^3 V_2^2 (3 \beta  U-z \beta  U'+1) \varphi ' z \Big)+ \\
\notag && V_2 \delta V_2 \Big(k^2 \beta  V_1^2 V_2 \varphi ^2 U' z^3+k^2 \beta  U V_1^2 \varphi ^2 V_2' z^3-V_1^3 V_2 U' (z^2 \beta  U \varphi '^2-1) z-U V_1^3 V_2' (z^2 \beta  U \varphi '^2-1) z - 2 z^3 V_1^3 V_2 A_{\tau} A_{\tau}' \Big)+ \\
\notag && V_2 \delta U \Big( 3 \beta  U V_1^2 V_2^2 V_1' \varphi '^2 z^3+k^2 \beta  V_1 V_2^2 \varphi ^2 V_1' z^3-k^2 \beta  V_1^2 V_2 \varphi ^2 V_2' z^3-V_1^2 V_2^2 (4 k^2 \beta  \varphi  \varphi ' z^2+V_1') z-  \\
\notag &&  V_1^3 V_2 V_2' (1-3 z^2 \beta  U \varphi '^2) z-4 V_1^3 V_2^2 (3 z^2 \beta  U \varphi '^2-1) \Big)+V_2 \delta V_2' \Big(-2 k^2 \beta  U V_1^2 V_2 \varphi ^2 z^3-2 U V_1^3 V_2 (1-z^2 \beta  U \varphi '^2) z \Big) \Bigg] \\
&& +\hat{\beta} \sigma \sqrt{V_1 V_2} A_{\tau} \delta A_{\tau}' \,.
\end{eqnarray}
In order to determine the surface term, we take into account the behavior of the fields at the horizon $z\sim 1$
\begin{eqnarray}
\notag && U(z) \sim  U'(1) (z-1) + \mathcal{O}((z-1)^2) \, ,  \\
\notag && V_1(z) \sim V_{1H}+ \mathcal{O}((z-1)) \, ,  \\
\notag && V_2(z) \sim V_{2H}+ \mathcal{O} ((z-1)) \, ,  \\
\notag && A_{\tau} (z) \sim A_{\tau}'(1) (z-1)+ \mathcal{O} ((z-1)^2) \, ,   \\
&& \varphi(z) \sim \varphi_{H} +\mathcal{O}((z-1)) \, .
\end{eqnarray}
and also the asymptotical behavior of the fields at infinity, which is given by
\begin{eqnarray}\label{eq-secVboundy}
\notag && U(z) \sim 1-(1+5 \beta) \lambda^2 z^2 +(M+(1+5\beta) \lambda^2) z^3+ \mathcal{O} (z^4) \, ,  \\
\notag && V_1(z) \sim 1-(1+5 \beta) \lambda^2 z^2+\alpha z^3 + \mathcal{O}(z^4) \, ,  \\
\notag && V_2(z) \sim 1-(1+5 \beta) \lambda^2 z^2 -\left( 8 \eta \lambda (1+6\beta)/3+\alpha  \right) z^3 + \mathcal{O}(z^4) \, ,   \\
\notag && A_{\tau} (z) \sim \mu-(\rho+\mu)z+ \mathcal{O}(z^2) \, ,  \\
&& \varphi(z) \sim \lambda z+ \eta z^2 + \mathcal{O}(z^3) \, ,
\end{eqnarray}
where, we have considered $\Lambda=-3$ as in the previous section. Then, the variation of the surface term at the horizon is
\begin{eqnarray}
\notag \delta B_{surf} |_{z=1}&=& -\hat{\beta} \sigma \Bigg( \frac{U'(1)V_{2H}(V_{1H}-k^2 \beta \varphi_{H}^2)}{4 V_{1H} \sqrt{V_{1H}V_{2H}}}  \delta V_{1H}+\frac{U'(1) (V_{1H}+k^2\beta \varphi_{H}^2)}{4 \sqrt{V_{1H} V_{2H}}} \delta V_{2H} \\
\notag && +\frac{U'(1) k^2 \beta V_{2H} \varphi_{H}}{\sqrt{V_{1H} V_{2H}}}  \delta \varphi_{H} \Bigg) \\
 &=& \delta \left( \sigma \frac{2\pi (V_{1H}+k^2 \beta \varphi_{H}^2) V_{2H}}{\sqrt{V_{1H} V_{2H}}} \right) \, ,
\end{eqnarray}
where we have used $\hat{\beta}=1/T=-4 \pi /U'(1)$.
From the above expression we obtain the following surface term at the horizon

\begin{equation}
B_{surf} |_{z=1}=\sigma \frac{2 \pi (V_{1H}+k^2 \beta \varphi_{H}^2)V_{2H}}{\sqrt{V_{1H} V_{2H}}} \, ,
\end{equation}

On the other hand, the variation of the surface term at infinity is given by
\begin{eqnarray} \label{asin}
\notag \delta B_{surf} |_{z \sim 0} & \approx & -\hat{\beta} \sigma \Bigg[ \left( -\frac{4 \beta \lambda}{z}-8 \beta \eta+ \mathcal{O}(z) \right) \delta \varphi'|_{z \sim 0} +\left( -\frac{2 \lambda (1+3 \beta)}{z^2}-\frac{4 \eta (1+3\beta)}{z} +\mathcal{O}(1) \right) \delta \varphi|_{z \sim 0} \\
 &+& \left( -\frac{1}{2z^2} +\mathcal{O}(1) \right) \delta V_1'|_{z \sim 0} +\left( -\frac{1}{2z^2}+ \mathcal{O}(1)\right) \delta V_2' |_{z \sim 0} +  \left( \frac{1}{z^3} +\mathcal{O}(z^{-1}) \right) \delta U |_{z \sim 0} - A_{\tau} \delta A_{\tau}' |_{z \sim 0} \Bigg] \,,
\end{eqnarray}
where
\begin{eqnarray}
\notag \delta \varphi' |_{z \sim 0} & \approx & \delta \lambda +2 \delta \eta z  + \mathcal{O}(z^2)\, ,   \\
\notag \delta V_1' |_{z \sim 0} & \approx & -4 (1+5 \beta) \lambda \delta \lambda z + 3 \delta \alpha z^2 + \mathcal{O} (z^3) \, , \\
\notag \delta \varphi |_{z \sim 0} & \approx & \delta \lambda z +\delta \eta z^2 + \mathcal{O} (z^3) \,, \\
\notag \delta V_2' |_{z \sim 0} & \approx & -4 (1+5 \beta) \lambda \delta \lambda z - \left(  8(1+6\beta) \delta (\lambda \eta)+3 \delta \alpha \right) z^2 +\mathcal{O} (z^3) \, ,\\
\notag \delta U |_{z \sim 0} & \approx & -2(1+5 \beta) \lambda \delta \lambda z^2 +\left( \delta M +2(1+5 \beta) \lambda \delta \lambda \right)  z^3 + \mathcal{O} (z^4) \, , \\
\delta A_{\tau}'|_{z \sim 0} & \approx &  - \delta (\rho +\mu) + \mathcal{O} (z) \, .
\end{eqnarray}
Notice that in (\ref{asin}) the divergent terms coming from the purely gravitational contribution exactly cancels the divergent terms coming from the scalar field, yielding a finite expression for the variation of the boundary term at infinity
\begin{equation}
\delta B_{surf} |_{z=0}= \hat{\beta} \sigma \left( -\delta M-4(1+6 \beta) \delta (\lambda \eta) -(1+5 \beta) \delta \lambda^2 +2(1+7\beta ) \lambda \delta \eta + 4(1+5\beta) \eta \delta \lambda - \Psi (\rho+\mu) \right) \, ,
\end{equation}
where the chemical potential is given by $\Psi= A_{\tau} (0)-A_{\tau}(1)=\mu$. It is necessary to impose boundary conditions on the scalar field, $\eta=\eta(\lambda)$, in order to remove the variations in the above equation. Considering $\eta= dW(\lambda) /d\lambda$, the surface term at infinity is
\begin{equation}
B_{surf} |_{z=0} =\hat{\beta} \sigma \left( -M-2(1+5 \beta) \lambda \frac{dW(\lambda)}{d \lambda} -(1+5 \beta)\lambda^2 +2(1+3 \beta) W(\lambda) -\Psi (\rho+\mu)\right) \, .
\end{equation}
Working in the grand canonical ensemble, we can fix the temperature and the chemical potential $\Psi$ at the horizon.
Therefore, the on-shell Euclidean action is given by
\begin{eqnarray}
I_{E} &=& B_{surf} |_{z=0}-B_{surf} |_{z=1} \\
\notag &=& -\sigma \frac{2 \pi (V_{1H}+k^2 \beta \varphi_{H}^2)V_{2H}}{\sqrt{V_{1H} V_{2H}}}+ \hat{\beta} \sigma \left(  -M-2(1+5 \beta) \lambda \frac{dW(\lambda)}{d \lambda} -(1+5 \beta)\lambda^2 +2(1+3 \beta) W(\lambda) -\Psi (\rho+\mu)\right)
 \, .
\end{eqnarray}
Using the above expression, we deduce the thermodynamics quantities: mass, electric charge and entropy, which are given respectively by
\begin{eqnarray}
 \notag && \mathcal{M}=\left( \frac{\partial}{\partial \hat{\beta}} -\frac{1}{\hat{\beta}} \Psi \frac{\partial}{ \partial \Psi}\right) I_{E} =\sigma \left(  -M-2(1+5 \beta) \lambda \frac{dW(\lambda)}{d \lambda} -(1+5 \beta)\lambda^2 +2(1+3 \beta) W(\lambda) \right) \, , \\
 \notag && \mathcal{Q}= -\frac{1}{\hat{\beta}} \frac{\partial I_{E}}{\partial \Psi}= \sigma (\rho +\mu)   \, , \\
  && S= \hat{\beta} \frac{\partial I_{E}}{\partial \hat{\beta}}-I_{E} =\frac{2 \pi \sigma (V_{1H}+k^2 \beta \phi_{H}^2)V_{2H}}{\sqrt{V_{1H}V_{2H}}} \,    .
\end{eqnarray}
The condition of positivity of the entropy implies $V_{1H}+k^2 \beta \phi_H^2>0$. The above expression for the mass requires to determine the boundary conditions at spacelike infinity. This means that $\eta$ ($=\varphi_+$) should generally be some function of $\lambda$ ($=\varphi_-$).
As we are interested in holographic applications, we demand boundary conditions for the scalar field to preserve the asymptotic AdS symmetry. The boundary conditions that the solution satisfies is determined by demanding that the solution is regular. This in turn determines what is the boundary condition that the hairy black holes are compatible with\cite{Papadimitriou:2007sj, Caldarelli:2016nni}.
A discussion about the boundary conditions for computing the mass of hairy black holes that preserve the asymptotic anti-de Sitter invariance  is given in \cite{Anabalon:2014fla}. In \cite{Witten:2001ua,Papadimitriou:2007sj} a general and systematic method to address multi-trace deformations and to properly account for the fact that the boundary is a conformal boundary was presented. We note that when the mass of the scalar field lies in the range $-\frac{5}{4}(1+3\beta)\geq m^2\geq -\frac{9}{4}(1+3\beta)$, the scalar hair satisfies mixed boundary conditions. The boundary conditions in our numerical hairy black hole solution deserve further studies.

Then, the free energy of the thermal sector is
\begin{equation}
\mathcal{F}=I_{E}/\hat{\beta}=\mathcal{M}-S/\hat{\beta}-\Psi \mathcal{Q}.
\end{equation}
Even though we have discussed the instability of system and the critical temperatures in last section, it would very interesting to evaluate the free energy to further figure out the phase structure of the solutions with possible boundary conditions. Due to the complexity of the calculations, we expect to study this question in the near future.

\subsection{Conductivity}

Having the fully backreacted  solution we will  calculate the conductivity and compare the results with the  conductivities
we found in the probe limit in section \ref{sec-HS}. To calculate the conductivity at the linear level  we consider the following  perturbations
\begin{equation}
\delta g_{tx}=\frac{1}{z^2}e^{-i \omega t} h_{tx} (z), \,\,\, \delta{\varphi }=i e^{i k x-i\omega t} z \delta \tilde{\varphi} (z), \,\,\, \delta A_x=e^{-i \omega t} a_x (z)~.
\end{equation}
We will consider two cases $k=0$ and $k\neq 0$ depending on having  momentum relaxation or not.

\subsubsection{$k=0$}

 In this case, the equation of the scalar field perturbation decoupled from the other equations. The coupled  differential equations for the electromagnetic and gravitational perturbations are
\begin{eqnarray}
  a_x''+\frac{a_x' \left(V_1 \left(2 V_2 U'+U V_2'\right)-U V_2 V_1'\right)}{2 U V_1 V_2}+ \frac{V_1 V_2 A_t' h_{tx}'-h_{tx} V_2 A_t' V_1'}{U V_1 V_2} + \frac{\omega ^2 a_x}{U^2} =0~,\\
  2 i \omega  z^2 a_x V_1 A_t'-i \omega  V_1 h_{tx}' \left(\beta  z^2 U \chi '^2-1\right)+i \omega  h_{tx} V_1' \left(\beta  z^2 U \chi '^2-1\right) =0~,
\end{eqnarray}
which  control the conductivity. Eliminating $h_{tx}$ from the above equation, we get
\begin{equation}\label{eq-perAx}
2 a_x V_1 V_2 \left(\frac{2 z^2 A_t'^2}{\beta  z^2 U \chi '^2-1}+\frac{\omega ^2}{U}\right)+2 U V_1 V_2 a_x''+a_x' \left(V_1 \left(2 V_2 U'+U V_2'\right)-U V_2 V_1'\right) =0\,.
\end{equation}

\begin{figure}[h]
\begin{center}
\includegraphics[width=70mm]{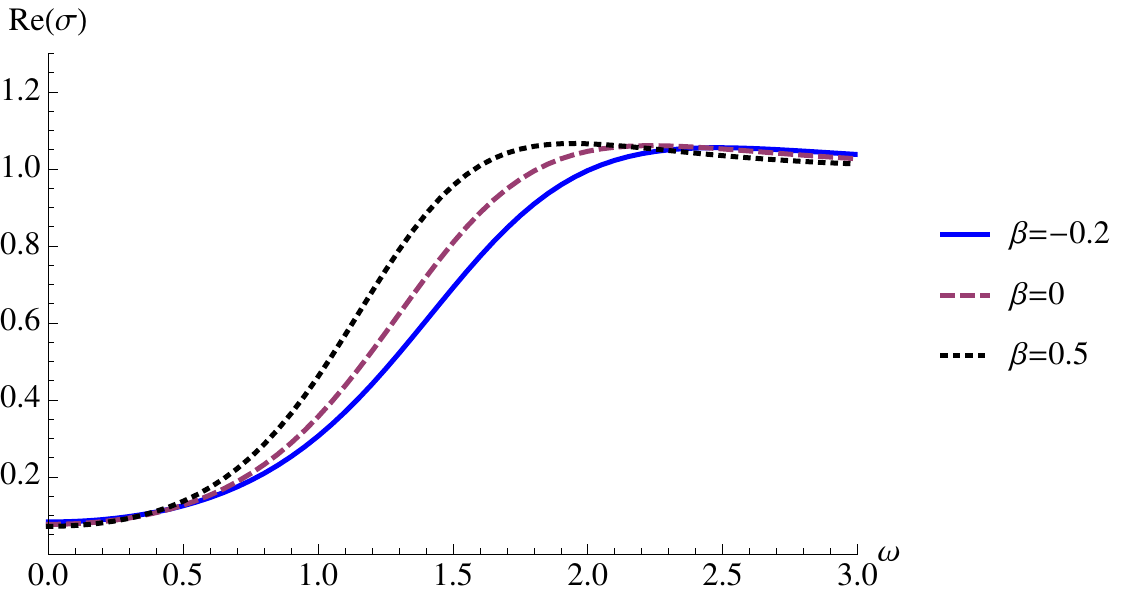}\hspace{1cm}
\includegraphics[width=70mm]{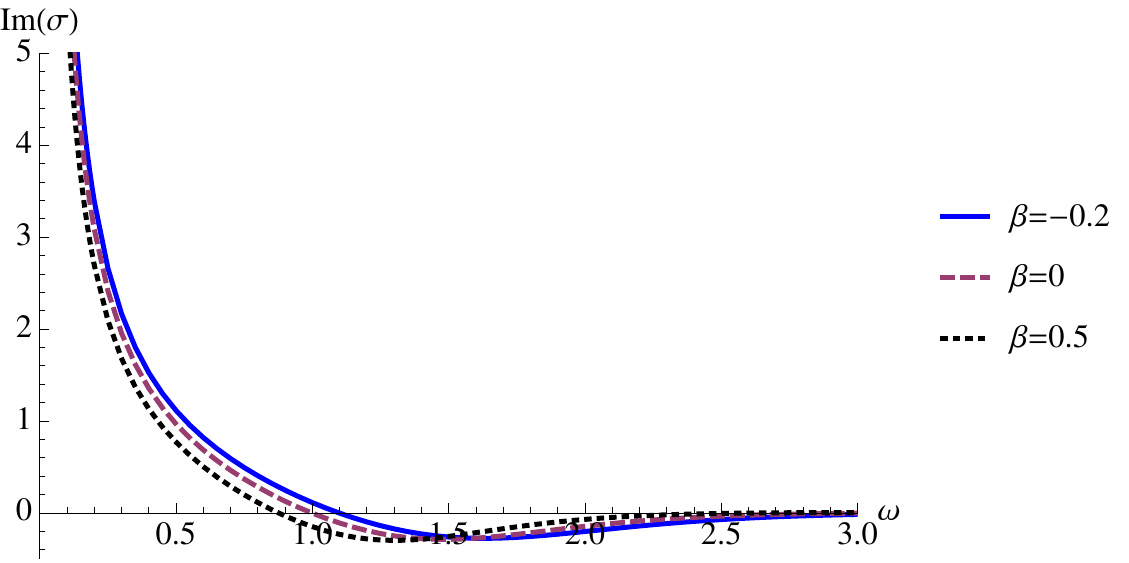}
\end{center}
\caption{The real and imaginary part of the conductivity $\sigma (\omega)$ as a function of $\omega$ with different couplings for $k / \mu = 0$. We set  $\lambda / \mu =0.3$ and $T / \mu = 1$.}
\label{fig-sigma-k0}
\end{figure}

When $z \rightarrow 0$, the asymptotic behavior of the perturbation is derived to be  $a_x (z)= a_x^0+ z a_x^1+ \mathcal{O} (z^2)$. Then, according to holographic dictionary,  the conductivity is given by
\begin{equation}\label{eq-Boundary-ax}
\sigma (\omega) = \frac{1}{i \omega}\frac{a_x^1}{a_x^0 }~.
\end{equation}
Near the black hole horizon $z \rightarrow 1$, we  impose purely ingoing boundary conditions
\begin{equation}
a_x (z)= (1-z^2)^{-i \omega / u(1)} ( a_x^H+\mathcal{O}(1-z))
\end{equation}
with $u(1)=-U'(1)=4 \pi T$ . Then, using the results from the last  subsection, we  solve equation \eqref{eq-perAx} at the boundary and calculate the conductivity using the relation \eqref{eq-Boundary-ax}.  The conductivity $\sigma(\omega)$ with $k=0$ is shown in Fig.~\ref{fig-sigma-k0}. In this case, the hairy solution is homogeneous
in spatial directions and the translational symmetry is hold. So the conductivity behaves similarly as that in RN black hole, the DC conductivity has a delta function at zero frequency due to the infinity of imaginary part and the real part of $\sigma$ approaches to $1$  at large frequency limit. Observe
that we have the same  behavior of the conductivity for both signs of the derivative coupling $\beta$.

\subsubsection{$k \neq 0$}\label{sec-backreaction22}

Now, we turn to the case with $k\neq 0$ in which the coupled system is more complicated because  of the highly coupled three perturbed equations, which are listed in the Appendix B.

To read off the conductivity, we analyze the boundary conditions of the perturbed fields.
The asymptotic behavior of the fields are $h_{tx} (z)=h_{tx}^0+ \mathcal{O}(z)$, $a_x (z)= a_x^0+ z a_x^1+ \mathcal{O} (z^2)$ and  $\delta \tilde{\varphi}(z)= \delta \tilde{\varphi}^0 + z \delta \tilde{\varphi}^1 +\mathcal{O}(z^2)$ and the expression of the  conductivity is also given in \eqref{eq-Boundary-ax}. The behavior of the fields near the horizon is
\begin{eqnarray}
a_x(z) &=& (1-z^2)^{-i \omega /u(1)}( \hat{a}_x^H+\mathcal{O}(1-z))~, \notag\\
h_{tx}(z) &=& (1-z^2)^{-i \omega /u(1)}( \hat{h}_{tx}^H (1-z)+\mathcal{O}((1-z)^2))~,\notag \\
\delta \tilde{\varphi} (z) &=& \delta \hat{\varphi}^H+\mathcal{O}(1-z)~,
\end{eqnarray}
with $u(1)=-U'(1)=4 \pi T$  which are all regular.

\begin{figure}[!h]
\begin{center}
\includegraphics[width=70mm]{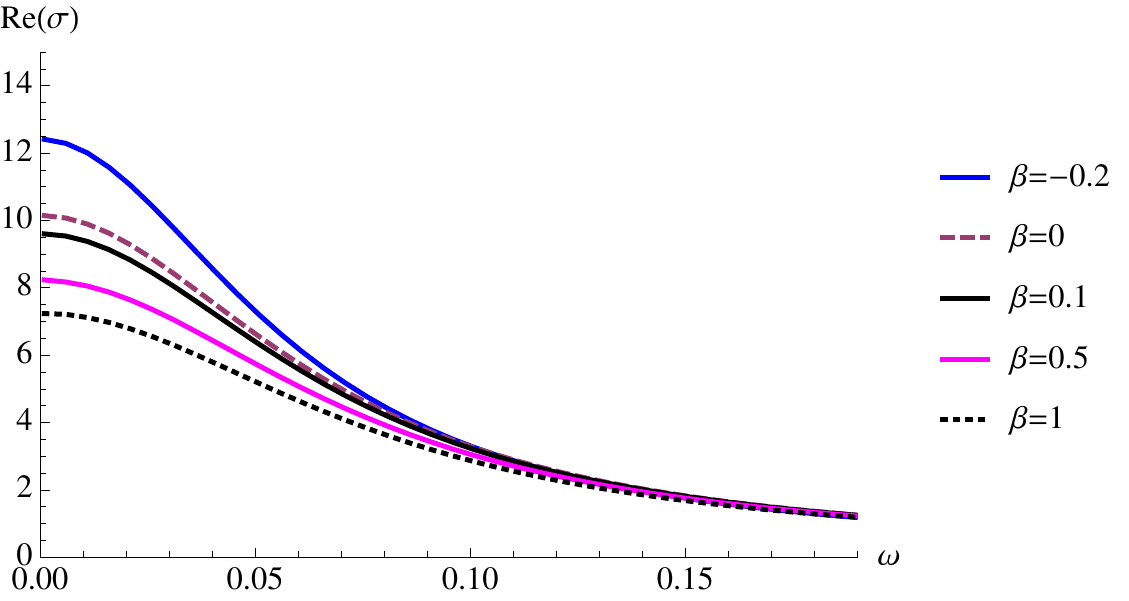}\hspace{1cm}
\includegraphics[width=70mm]{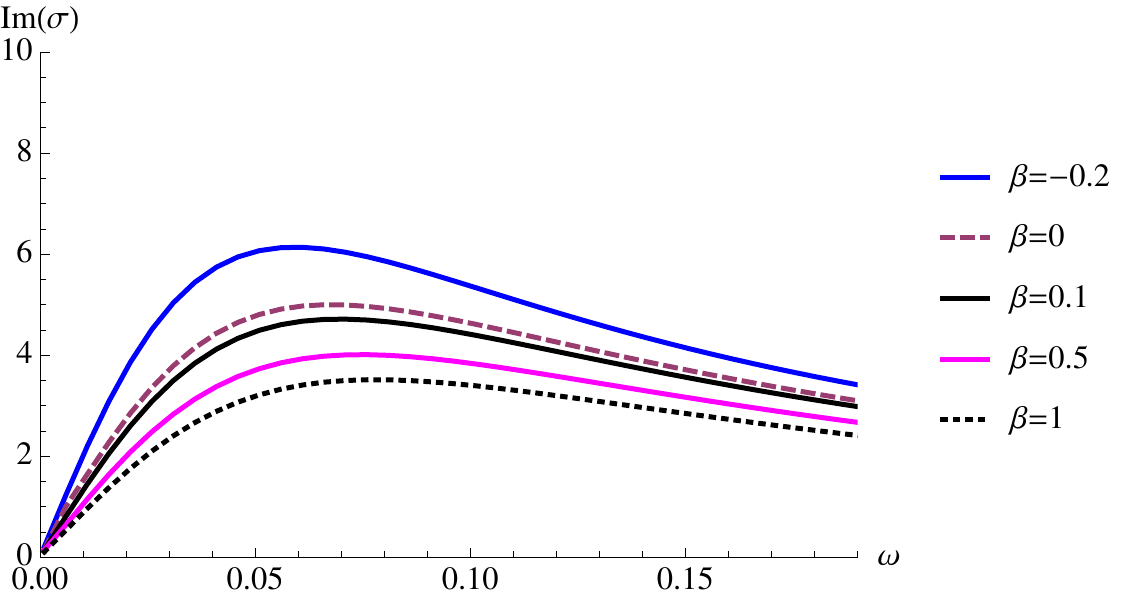}
\end{center}
\caption{The real and imaginary part of the conductivity $\sigma (\omega)$ with different couplings for  $k/\mu=1/\sqrt{2}$. We set $\lambda/\mu=0.5$ and $T/\mu=0.1$.}
\label{fig-sigma-k}
\end{figure}

The conductivity for different couplings $\beta$ is shown in Fig. \ref{fig-sigma-k} with fixed $k/\mu=1/\sqrt{2}$, $\lambda/\mu=0.5$ and $T/\mu=0.1$. We see with nonzero $k$, the DC conductivity $\sigma(0)$ is finite which according to the holographic dictionary our hairy black hole solution in the bulk is dual to a material with momentum relaxation on the boundary. If $\beta=0$ we recover the results of the Q-Lattice \cite{Donos:2013eha} in which the translational invariance is explicitly broken. We observe that as the derivative coupling  $\beta$ is increased in positive values the electric conductivity $\sigma(0)$ becomes  lower while if $\beta$  is negative, $\sigma(0)$ is enhanced.

 It is interesting to see how the DC conductivity varies with the temperature for various values of the coupling $\beta$. We expect in the dual theory at low temperatures $T<<\mu$,
  the DC conductivity to increase as the temperature is lower and the material to enter a metal phase, while as the temperature is increased the material to enter an insulating phase with the DC conductivity to decrease.

\begin{figure}[!h]
\begin{center}
\includegraphics[width=70mm]{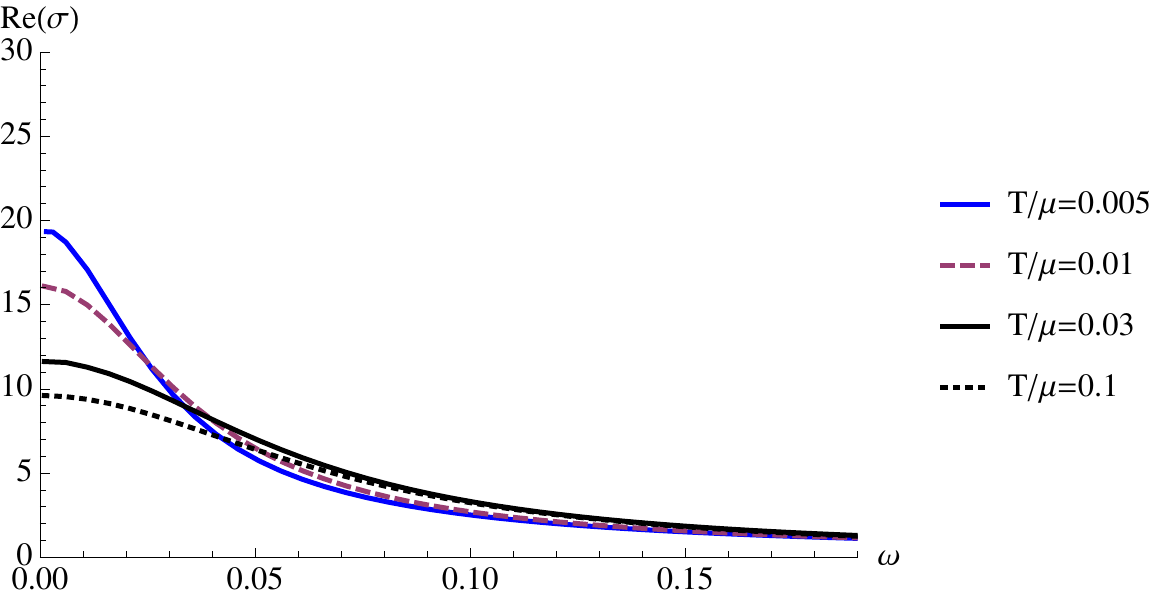}
\includegraphics[width=70mm]{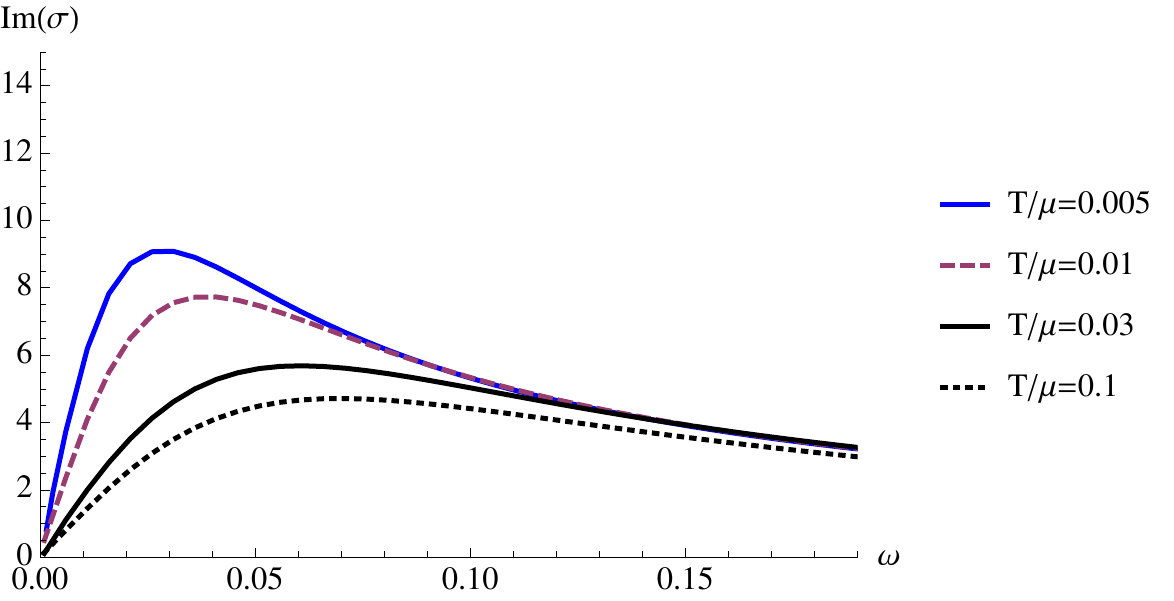}

\end{center}
\caption{The real and imaginary part of the conductivity $\sigma (\omega)$ for different temperatures with positive coupling $\beta=0.1$. We set $k/\mu=1/\sqrt{2}$ and $\lambda/\mu=0.5$. }
\label{fig-con-beta0p1}
\end{figure}

\begin{figure}[!h]
\begin{center}
\includegraphics[width=70mm]{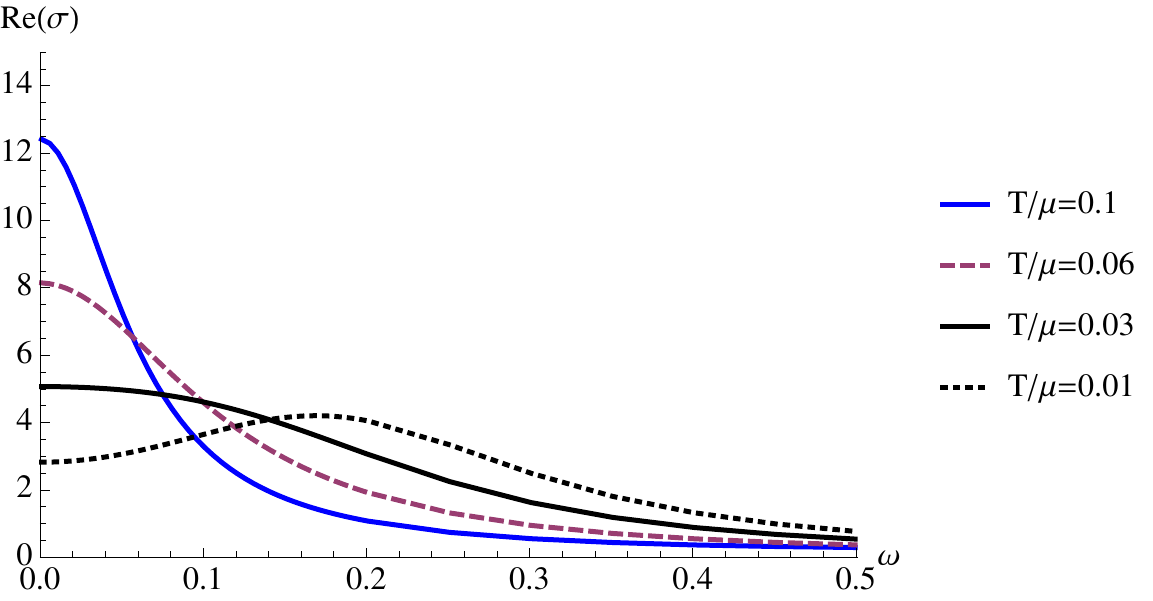}
\includegraphics[width=70mm]{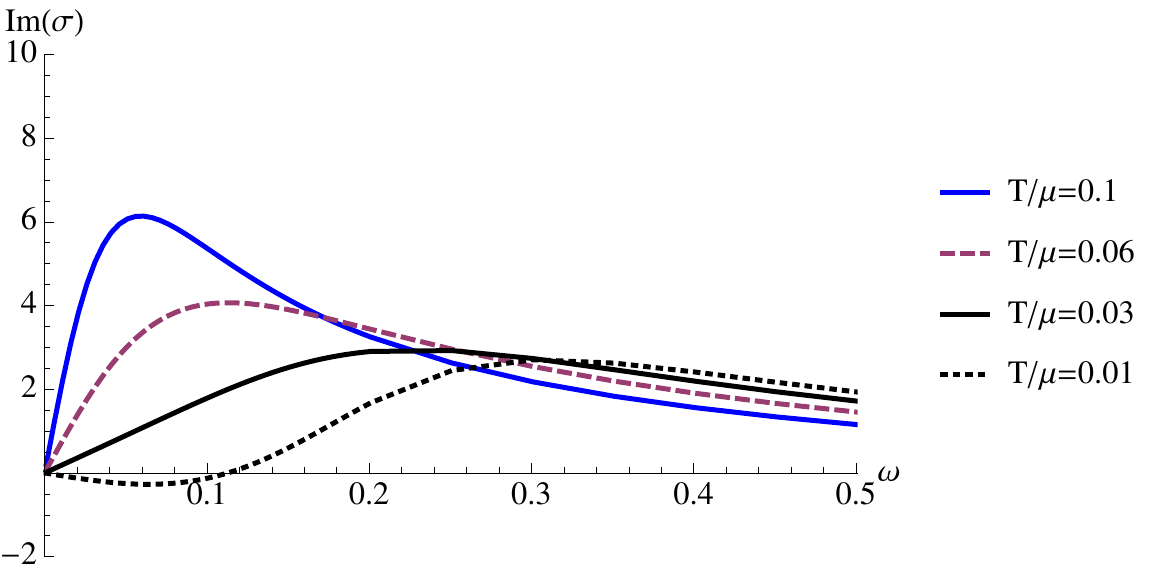}

\end{center}
\caption{The real and imaginary part of the conductivity $\sigma (\omega)$ for different temperatures with negative coupling  $\beta=-0.2$. We set  $k/\mu=1/\sqrt{2}$, $\lambda/\mu=0.5$.}
\label{fig-con-betam0p2}
\end{figure}

  We observe these  changes of phases as we vary the coupling $\beta$. In Fig.~\ref{fig-con-beta0p1} we fix $\beta=0.1$ and show the conductivities for different low enough temperatures. We see that the dual theory  is in a  metal phase  because DC conductivity increases as the temperature decreases. In Fig.~\ref{fig-con-betam0p2} for $\beta=-0.2$, the  DC conductivity decreases as we decreases $T$ and  the dual theory is in an  insulating phase.
  The Q-Lattice model in \cite{Donos:2013eha} also supports two phases of a doped material in the boundary theory. However, these two phases are dual to  two different black hole solutions in the bulk. In our case the coupling $\beta$ plays the role of the doping parameter.

\begin{figure}[!h]
\begin{center}
\includegraphics[width=70mm]{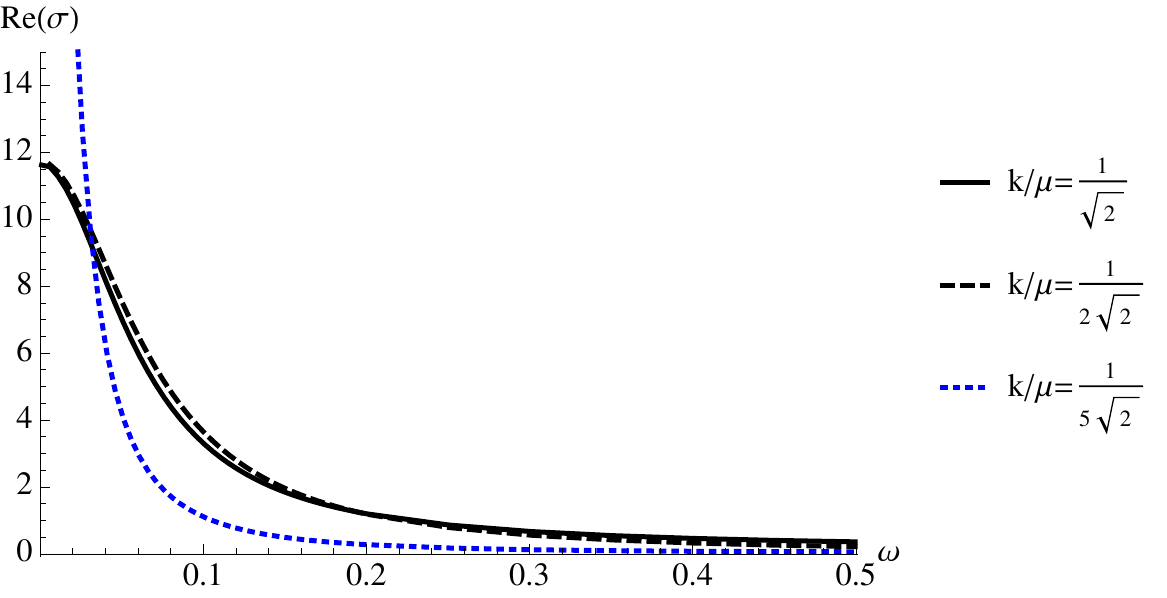}
\includegraphics[width=70mm]{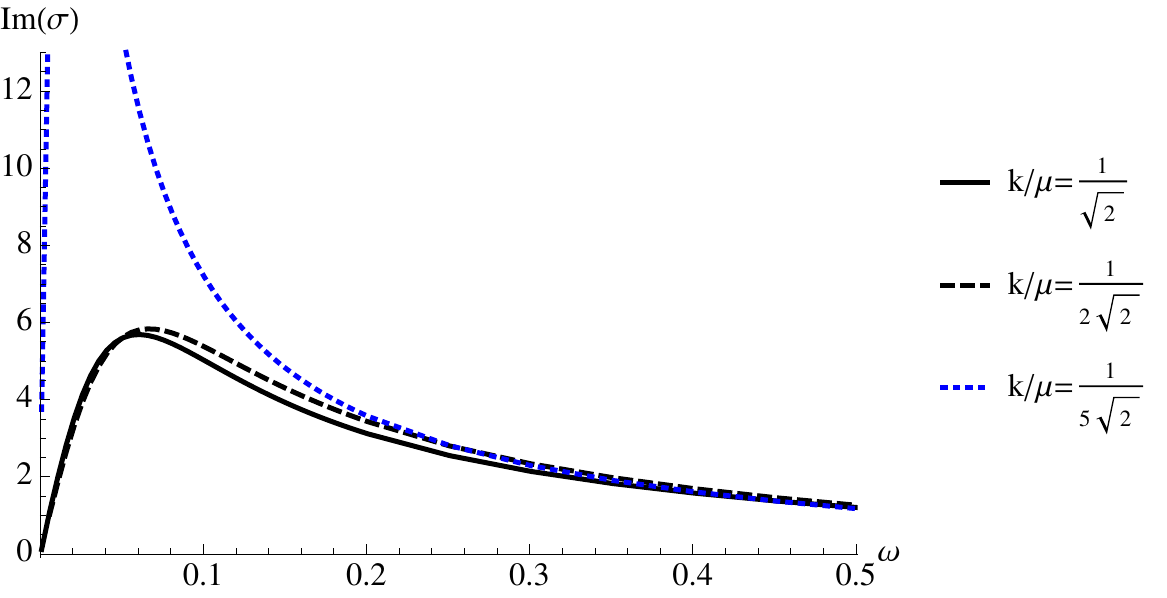}

\end{center}
\caption{The real and imaginary part of the conductivity $\sigma (\omega)$ for different k with positive coupling $\beta=0.1$. We set $T/\mu=0.03$ and $\lambda/\mu=0.5$. }
\label{fig-con-figp112}
\end{figure}

\begin{figure}[!h]
\begin{center}
\includegraphics[width=70mm]{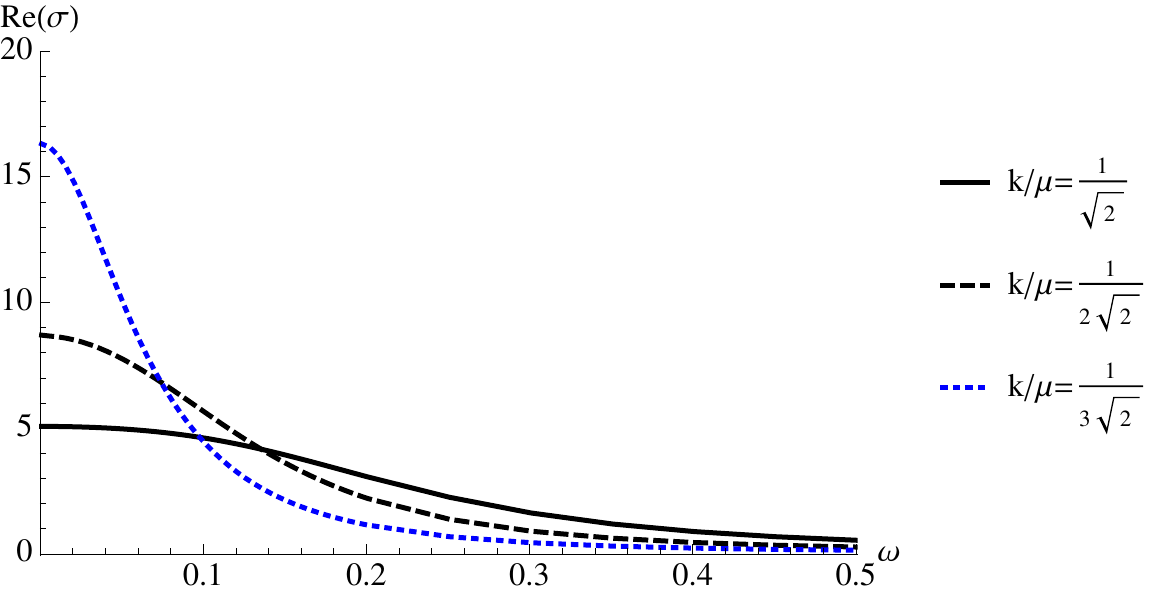}
\includegraphics[width=70mm]{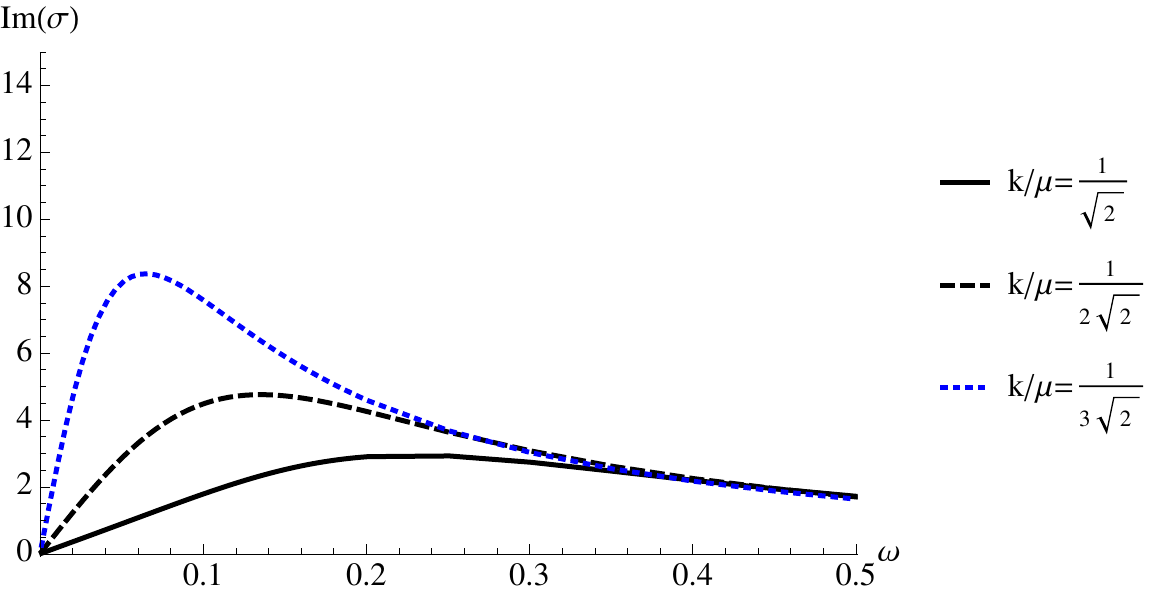}

\end{center}
\caption{The real and imaginary part of the conductivity $\sigma (\omega)$ for different k with negative coupling $\beta=-0.2$. We set $T/\mu=0.03$ and $\lambda/\mu=0.5$. }
\label{fig-con-figp134}
\end{figure}

 In Fig.~\ref{fig-con-figp112} and Fig.~\ref{fig-con-figp134} we show the conductivity for different momentum dissipation numbers $k$ for positive and negative coupling $\beta$ and for fixed low temperature. They show a competing effect between $k$ and the coupling  $\beta$. In the metal phase Fig.~\ref{fig-con-figp112} shows that if the momentum dissipation is large the conductivity is low. This can be understood from the fact that  the charge carriers for large momentum dissipation have more chances  of finding impurities, because of the presence of a non-zero coupling $\beta$, so they loose energy and for this reason the conductivity is low. In the insulating phase Fig.~\ref{fig-con-figp134} shows that even for large momentum dissipation the conductivity is low, because the charge carriers  do not have the freedom to travel.

  In \cite{Kuang:2016edj} it was
  shown that in the probe limit  the bulk theory is   dual to a material in a metal phase and the positive coupling $\beta$ signifies the amount of impurities in the material. In the fully  backreacted theory we study in this work if the coupling $\beta$ is positive we still have impurities  but their concentration is not high enough to prevent conductivity. However, if $\beta$ is negative then the friction on the velocities of the charge carriers is so high that the system  enters the insulating phase. A change of sign of the coupling $\beta$ in the gravity sector corresponds to a change of sign of the kinetic energy of the scalar field coupled to Einstein tensor. It is interesting to note that in the boundary theory  due to proximity effect \cite{Buzdin:2005zz} there is a reflection of the charge carriers on the interface between the normal and the superconducting phases of the material. This is known as the Andreev Reflection effect \cite{Pannetier}.
  Note also, this only happens  in very low temperatures which means that there is no thermal energy to influence the transition. This behavior of the coupling $\beta$  is a very interesting effect, it is a pure gravity effect because the derivative coupling is showing how strong is the coupling of matter to curvature.


\section{Conclusions}
\label{sec-conc}

In this paper, we  analyzed the possible dual phases in Horndeski theory with a coupling between the scalar field and the Einstein tensor. Extending  the holographic study in \cite{Kuang:2016edj}, we considered the simple inhomogeneous matter field  in the probe limit,  and studied the holographic superconducting phase transition and the conductivities  of the dual boundary theory. The AC conductivity shows a non-trivial structure indicating a collective excitation of the charge carriers as a result of the  breaking of translation invariance and the presence of the coupling of the scalar field to the Einstein tensor.

We then discussed possible instabilities in the theory and we analyzed the spontaneous breaking of translation invariance  near the critical temperature. We studied the fully backreacted system of  Einstein-Maxwell-scalar equations and numerically found the hairy black hole solution, and then we also analyze the thermodynamic of the hairy black hole solution. We  computed the conductivity of the dual theory and studied the generated phase. For the zero wave number of the scalar field, the DC conductivity was divergent as expected and  the system is  dual to ideal conductor. For nonzero wave number, the  DC conductivity was finite having momentum dissipation on the dual boundary theory.  At low temperatures, we found that for positive coupling the DC conductivity  increases as  the temperature is lower, indicating that its dual phase is a metal. For negative coupling we found  the DC conductivity to decrease as  the temperature is lower indicating that  the dual phase is an insulator.

Our results are interesting and deserve a further study. We found that there is a change of a phase in the boundary theory as we change the value and the sign of the
coupling of a charged scalar field to the Einstein tensor. This coupling shows the way matter is coupled to curvature and it is a pure gravity effect.
On the dual theory our results show that the variation of this coupling influences the kinetic properties of the charge carriers. In a way this coupling parameterizes the amount of impurities present in a material on the boundary. On the other hand, a change on the sign of the kinetic energy of the scalar field allows the transition from one phase to an another in the boundary theory.

\acknowledgments

We thank Matteo Baggioli for participating in the first stages of this work and for numerous discussions we had with him.  We also thank the referee whose comments and remarks have improved the
quality of this manuscript significantly.
X.M Kuang is supported by the Natural Science Foundation of China under Grant No.11705161 and Natural Science Foundation of Jiangsu Province under Grant No.BK20170481. Y. V. acknowledges to the organizing committee of the school on numerical methods in gravity and holography, 2017, held in the Universidad de Concepci\'on, for financial support. This work was funded by the Direcci\'{o}n de Investigaci\'{o}n y Desarrollo de la Universidad de La Serena (Y.V.).


\section*{Appendix A: Independent equations of
motions with full backreaction}
In this appendix, we show the equations of motion we solve to determine the hairy black hole solution in section \ref{sec-backreaction1}.
The Einstein field equations are
\begin{itemize}
\item tt component
\begin{align}
\nonumber &0= -V_1^3 \Big(2 V_2^2 \left(z^4 A_t'^2+2 \Lambda +2 m^2 \varphi^2+2 U \left(z^2 \varphi ' \left(\varphi ' \left(3 \beta  z U'+\beta  U+1\right)+4 \beta  z U \varphi ''\right)+3\right)-2 z U'\right) \\
\nonumber & +z V_2 \left(z U' V_2'+U \left(2 z V_2''-V_2' \left(3 \beta  z^3 U' \varphi '^2+4\right)\right)-2 \beta  z^3 U^2 \varphi ' \left(V_2'' \varphi '+2 V_2' \varphi ''\right)\right)+z^2 U V_2'^2 \left(\beta  z^2 U \varphi '^2-1\right)\Big)  \\
\nonumber  & +z^2 V_1 V_2 \Big(V_1' \left(\beta  k^2 z^2 V_2 \varphi^2 U'+U \left(V_2 \left(4 \beta  k^2 z \varphi \left(2 z \varphi '+\varphi\right)+V_1'\right)+\beta  k^2 z^2 \varphi^2 V_2'\right)-\beta  z^2 U^2 V_2 V_1' \varphi '^2\right)+ \\
\nonumber &  2 \beta  k^2 z^2 U V_2 \varphi^2 V_1''\Big)-z V_1^2 \Big(V_2 \left(\beta  k^2 z^3 \varphi^2 \left(U' V_2'+2 U V_2''\right)+4 \beta  k^2 z^3 U \varphi V_2' \varphi '+z U V_1' V_2' \left(1-\beta  z^2 U \varphi '^2\right)\right)  \\
\nonumber & +V_2^2 \Big(U \left(8 \beta  k^2 z^3 \varphi '^2-V_1' \left(3 \beta  z^3 U' \varphi '^2+4\right)+2 z V_1''\right)+4 \beta  k^2 z^2 \varphi \left(\left(z U'+4 U\right) \varphi '+2 z U \varphi ''\right)  \\
 & +4 k^2 z \varphi^2 (\beta  U+1)+z U' V_1'-2 \beta  z^3 U^2 \varphi ' \left(V_1'' \varphi '+2 V_1' \varphi ''\right)\Big)+\beta  \left(-k^2\right) z^3 U \varphi^2 V_2'^2\Big)-3 \beta  k^2 z^4 U V_2^2 \varphi^2 V_1'^2
\end{align}

\item zz component
\begin{align}
\nonumber &0= V_1^2 \Big(2 V_2 \left(z^4 A_t'^2+2 \Lambda +2 m^2 \varphi^2+2 U \left(z^2 \varphi '^2 \left(3 \beta  z U'-9 \beta  U-1\right)+3\right)-2 z U'\right) \\
\nonumber & +z \left(z U'-4 U\right) V_2' \left(1-3 \beta  z^2 U \varphi '^2\right)\Big)+\beta  \left(-k^2\right) z^3 \varphi^2 V_1' \left(z V_2 U'+U \left(z V_2'-4 V_2\right)\right) \\
\nonumber & +z V_1 \Big(z V_2' \left(\beta  z^2 \left(k^2 \varphi^2 U'+4 k^2 U \varphi \varphi '-3 U^2 V_1' \varphi '^2\right)+U V_1'\right)+V_2 \Big(4 \beta  k^2 z^2 \varphi \left(z U'-4 U\right) \varphi ' \\
& -4 k^2 z \varphi^2 (\beta  U-1)+\left(4 U-z U'\right) V_1' \left(3 \beta  z^2 U \varphi '^2-1\right)\Big)\Big)
\end{align}

\item xx component
\begin{align}
\nonumber&0=  V_1 \Big(2 V_2^2 \Big(-z \left(z^3 \left(A_t'^2+\beta  U'^2 \varphi '^2\right)-z U''+4 U'\right)+2 \Lambda +2 m^2 \varphi^2+U \Big(z^2 \varphi ' \Big(\varphi ' \left(\beta  z \left(6 U'-z U''\right)+2\right) \\
\nonumber & -2 \beta  z^2 U' \varphi ''\Big)+6\Big)+2 \beta  z^2 U^2 \varphi ' \left(4 z \varphi ''+\varphi '\right)\Big)-2 z V_2 \Big(-z U' V_2'+U \left(2 V_2' \left(\beta  z^3 U' \varphi '^2+1\right)-z V_2''\right)  \\
\nonumber & +\beta  z^3 U^2 \varphi ' \left(V_2'' \varphi '+2 V_2' \varphi ''\right)\Big)+z^2 U V_2'^2 \left(\beta  z^2 U \varphi '^2-1\right)\Big)+k^2 z^2 \varphi^2 \Big(-2 \beta  z V_2 \left(\left(z U'-2 U\right) V_2'+z U V_2''\right) \\
&-2 V_2^2 \left(\beta  z \left(z U''-4 U'\right)+6 \beta  U+2\right)+\beta  z^2 U V_2'^2\Big)
\end{align}

\item yy component
\begin{align}
\nonumber &0= 2 V_1^3 \Big(-z \left(z^3 \left(A_t'^2+\beta  U'^2 \varphi '^2\right)-z U''+4 U'\right)+2 \Lambda +2 m^2 \varphi^2+U \Big(-2 \beta  z^4 U' \varphi ' \varphi ''+z^2 \varphi '^2 \Big(\beta  z (6 U' \\
\nonumber &-z U'')+2\Big)+6\Big)+2 \beta  z^2 U^2 \varphi ' \left(4 z \varphi ''+\varphi '\right)\Big)+z^2 V_1 \Big(-V_1' \Big(2 \beta  k^2 z^2 \varphi^2 U'+U \left(4 \beta  k^2 z \varphi \left(2 z \varphi '+\varphi\right)+V_1'\right)   \\
\nonumber &-\beta  z^2 U^2 V_1' \varphi '^2\Big)-2 \beta  k^2 z^2 U \varphi^2 V_1''\Big)+2 z V_1^2 \Big(k^2 z \varphi^2 \left(\beta  z^2 U''+2 \beta  U+2\right)+U \Big(4 \beta  k^2 z^3 \varphi '^2 \\
\nonumber &  -2 V_1' \left(\beta  z^3 U' \varphi '^2+1\right)+z V_1''\Big)+4 \beta  k^2 z^2 \varphi \left(\left(z U'+2 U\right) \varphi '+z U \varphi ''\right) \\
&+z U' V_1'-\beta  z^3 U^2 \varphi ' \left(V_1'' \varphi '+2 V_1' \varphi ''\right)\Big)  +3 \beta  k^2 z^4 U \varphi^2 V_1'^2
\end{align}
\end{itemize}

The time component of the Maxwell Equations gives
\begin{equation} \label{maxwell}
-2 V_1 V_2 A_t''-A_t' \left(V_2 V_1'+V_1 V_2'\right) = 0
\end{equation}

and the Klein Gordon equation has the form
\begin{align}
\nonumber &0= z^2 V_1 \Big(2 \beta  z V_2 \left(U \left(2 k^2 \varphi \left(2 V_2'-z V_2''\right)+3 z U' V_1' V_2' \varphi '\right)-2 k^2 z \varphi U' V_2'+U^2 \left(z V_1' V_2' \varphi ''+\varphi ' \left(z V_1'' V_2'+V_1' \left(z V_2''-4 V_2'\right)\right)\right)\right) \\
\nonumber &+2 V_2^2 \Big(-2 k^2 \varphi \left(\beta  z \left(z U''-4 U'\right)+6 \beta  U+2\right)+\beta  z U \left(z U'-4 U\right) V_1' \varphi ''+\varphi ' \Big(\beta  z U \left(z U'-4 U\right) V_1''  \\
\nonumber & +V_1' \left(\beta  z^2 U'^2+U \left(\beta  z \left(z U''-10 U'\right)+2\right)+10 \beta  U^2\right)\Big)\Big)+\beta  z^2 U V_2'^2 \left(2 k^2 \varphi-U V_1' \varphi '\right)\Big) \\
\nonumber &+V_1^2 \Big(8 V_2^2 \Big(z \left(z U \varphi '' \left(-\beta  z U'+3 \beta  U+1\right)-\varphi ' \left(z U' \left(\beta  z U'-1\right)+U \left(\beta  z \left(z U''-7 U'\right)+2\right)+6 \beta  U^2\right)\right) \\
\nonumber &-m^2 \varphi\Big)+\beta  z^3 U \left(4 U-z U'\right) V_2'^2 \varphi '+2 z^2 V_2 \Big(\beta  z U \left(z U'-4 U\right) V_2' \varphi ''+\varphi ' \Big(\beta  z U \left(z U'-4 U\right) V_2'' \\
&  +V_2' \left(\beta  z^2 U'^2+U \left(\beta  z \left(z U''-10 U'\right)+2\right)+10 \beta  U^2\right)\Big)\Big)\Big)-\beta z^3 U V_2 V_1'^2 \varphi ' \left(z V_2 U'+U \left(z V_2'-4 V_2\right)\right).
\end{align}

\section*{Appendix B: Coupled perturbed equations determining the conductivity}
In this appendix, we show the perturbed equations of motion we solve to determine conductivity for $k\neq 0$ in section \ref{sec-backreaction22}.
The perturbed Maxwell equation is
\begin{align}
0=U^2 \left(-V_2\right) V_1' a_x'+U^2 V_1 V_2' a_x'+2 U^2 V_1 V_2 a_x''+2 U V_1 V_2 U' a_x'+2 V_1 V_2 \omega ^2 a_x+2 U V_1 V_2 A_t' h_{\text{tx}}'-2 U V_2 h_{\text{tx}} V_1' A_t',
\end{align}
the $tx-$coponent of Einstein equation for the perturbed $h_{tx}$ gives
\begin{align}
\nonumber &0=2 V_1^2 V_2 \omega  z^2 a_x A_t'+V_1 V_2 \Big(\beta  \left(-k^2\right) \omega  z^2 \varphi^2 h_{\text{tx}}'+\omega  h_{\text{tx}} V_1' \left(\beta  U z^2 \varphi '^2-2\right)
-i \beta  k U z \delta \tilde{\varphi} ' \varphi \left(z U' V_1'-2 U \left(V_1'-z V_1''\right)\right)\\ \nonumber &+i \beta  \delta \tilde{\varphi}  k U z \varphi ' \left(z U' V_1'+2 U z V_1''-2 U V_1'\right)\big)
+\beta  k V_2 z^2 V_1'\left(k \omega  h_{\text{tx}} \varphi^2+i U^2 V_1' \left(\delta \tilde{\varphi} ' \varphi-\delta \tilde{\varphi}  \varphi '\right)\right)\\  &-i V_1^2 \left(\beta  k U z V_2' \left(2 U-z U'\right) \left(\delta \tilde{\varphi} '\varphi-\delta \tilde{\varphi}  \varphi '\right)+V_2 \left(4 k U \left(\delta \tilde{\varphi}  \varphi '-\delta \tilde{\varphi} ' \varphi \right)-i \omega  h_{\text{tx}}' \left(\beta  U z^2 \varphi'^2-2\right)\right)\right)
\end{align}
and the perturbed Klein-Gordon equation is
\begin{align}
\nonumber&0=z \beta  V_1 \Big(-4 \Big(6 \left(2 \delta \tilde{\varphi} '-z \delta \tilde{\varphi} ''\right) V_2^2+z \left(V_2' \left(2 z \delta \tilde{\varphi} ''-5 \delta \tilde{\varphi} '\right)+2 z \delta \tilde{\varphi} ' V_2''\right) V_2-z^2 \delta \tilde{\varphi}
   ' \left(V_2'\right){}^2\Big) V_1^2\\ \nonumber &+z \left(4 \left(V_1' \left(5 \delta \tilde{\varphi} '-2 z \delta \tilde{\varphi} ''\right)-2 z \delta \tilde{\varphi} ' V_1''\right) V_2^2+2 z \left(z \delta \tilde{\varphi} ' V_2' V_1''+V_1'
   \left(V_2' \left(z \delta \tilde{\varphi} ''-4 \delta \tilde{\varphi} '\right)+z \delta \tilde{\varphi} ' V_2''\right)\right) V_2-z^2 \delta \tilde{\varphi} ' V_1' \left(V_2'\right){}^2\right) V_1\\ \nonumber &+z^2 V_2 \delta \tilde{\varphi} '
   \left(V_1'\right){}^2 \left(4 V_2-z V_2'\right)\Big) U^3+z V_1 \Big(-\beta  V_2^2 U' \delta \tilde{\varphi} ' \left(V_1'\right){}^2 z^3+2 V_1 \Big(\big(-12 \beta  \delta \tilde{\varphi}  k^2+V_1'
   \big(\beta  U' \delta \tilde{\varphi} '' z^2+\delta \tilde{\varphi} ' \big(\beta  U'' z^2\\ \nonumber &-10 \beta  U' z+2\big)\big)+z^2 \beta  U' \delta \tilde{\varphi} ' V_1''\big) V_2^2+z \beta  \left(2 \delta \tilde{\varphi}  \left(2 V_2'-z V_2''\right) k^2+3 z U' \delta \tilde{\varphi} ' V_1' V_2'\right) V_2+k^2 z^2 \beta  \delta \tilde{\varphi}  \left(V_2'\right){}^2\Big) z\\ \nonumber &+V_1^2 \Big(-\beta  U' \delta \tilde{\varphi} ' \left(V_2'\right){}^2 z^3+2 V_2 \left(\beta  U' \delta \tilde{\varphi} ' V_2'' z^2+V_2' \left(\beta  U' \delta \tilde{\varphi} '' z^2+\delta \tilde{\varphi} ' \left(\beta  U'' z^2-10 \beta  U' z+2\right)\right)\right) z\\ \nonumber &-8 V_2^2
   \left(\delta \tilde{\varphi} '\left(\beta  U'' z^2-7 \beta  U' z+2\right)+z \left(z \beta  U'-1\right) \delta \tilde{\varphi} ''\right)\Big)\Big) U^2-2 \big(-3 i k \beta  \omega  h_{\text{tx}} V_2^2 \varphi
   \left(V_1'\right){}^2 z^4+i \beta  \omega  V_1 V_2 \big(k h_{\text{tx}} \varphi V_1' V_2'\\ \nonumber &+V_2 \left(2 k \varphi \left(h_{\text{tx}}' V_1'+h_{\text{tx}} V_1''\right)+V_1' \left(4 k
   h_{\text{tx}} \varphi '-i \delta \tilde{\varphi}  \omega  V_1'\right)\right)\big) z^4+V_1^2 \big(\big(2 \delta \tilde{\varphi}  \big(-4 z \beta  U' k^2+z^2 \beta  U'' k^2+2 k^2+2 z \beta  \omega ^2 V_1'\\ \nonumber &-z^2 \beta  \omega ^2 V_1''\big)+i \beta  \left(12 k \omega  h_{\text{tx}} \varphi (z)+z \left(i z \delta \tilde{\varphi} ' V_1' \left(U'\right)^2-4 k \omega  \varphi h_{\text{tx}}'-4 k z \omega
   h_{\text{tx}}' \varphi '\right)\right)\big) V_2^2+z \beta  \big(z \delta \tilde{\varphi}  \left(2 k^2 U'-\omega ^2 V_1'\right) V_2'\\ \nonumber &-2 i k \omega  h_{\text{tx}} \varphi \left(2 V_2'-z
   V_2''\right)\big) V_2-i k z^2 \beta  \omega  h_{\text{tx}} \varphi \left(V_2'\right){}^2\big) z^2+V_1^3 \big(\beta  \delta \tilde{\varphi}  \omega ^2 \left(V_2'\right){}^2 z^4-\beta  V_2 \big(2
   \delta \tilde{\varphi}  \left(z V_2''-2 V_2'\right) \omega ^2\\ \nonumber &+z \left(U'\right)^2 \delta \tilde{\varphi} ' V_2'\big) z^3+4 V_2^2 \left(U' \left(z \beta  U'-1\right) \delta \tilde{\varphi} ' z^2+\delta \tilde{\varphi}
   \left(m^2-3 z^2 \beta  \omega ^2\right)\right)\big)\big) U\\  &+2 z^2 \omega  V_1 V_2 \left(\delta \tilde{\varphi}  \omega  V_1-i k h_{\text{tx}} \varphi\right) \left(\beta  V_2 U' V_1' z^2+V_1 \left(\beta  U' V_2' z^2+V_2 \left(4-4 z \beta  U'\right)\right)\right).
\end{align}

\end{document}